\documentclass[12pt, titlepage]{article}

\usepackage[toc,page]{appendix}
\usepackage{booktabs} 
\usepackage[center]{caption}
\usepackage{subcaption}
\usepackage{cite}
\usepackage[shortlabels]{enumitem}
\usepackage{fancyhdr}
\usepackage[margin=1in]{geometry}

\usepackage{graphicx} 
\usepackage{hyperref}
\usepackage{indentfirst}
\usepackage{longtable}
\usepackage{pdfpages}
\usepackage[raggedright]{titlesec}
\usepackage{url}
\usepackage{verbatim}
\usepackage{wrapfig}
\usepackage{comment}
\usepackage{multirow}
\usepackage{multicol}
\usepackage{tabularx}
\usepackage{colortbl}

\hypersetup{
    colorlinks=true,
    linkcolor=blue,
    filecolor=magenta,      
    urlcolor=blue,
    citecolor =blue,
    pdfpagemode=FullScreen,
}

\makeatletter
\renewcommand\paragraph{\@startsection{paragraph}{4}{\z@}%
            {-2.5ex\@plus -1ex \@minus -.25ex}%
            {1.25ex \@plus .25ex}%
            {\normalfont\normalsize\bfseries}}
\makeatother

\titleformat{\paragraph}
{\normalfont\normalsize\bfseries}{\theparagraph}{1em}{}
\titlespacing*{\paragraph}
{0pt}{3.25ex plus 1ex minus .2ex}{1.5ex plus .2ex}

{\fontsize{10}{12}
\rhead{\includegraphics[width=4cm]{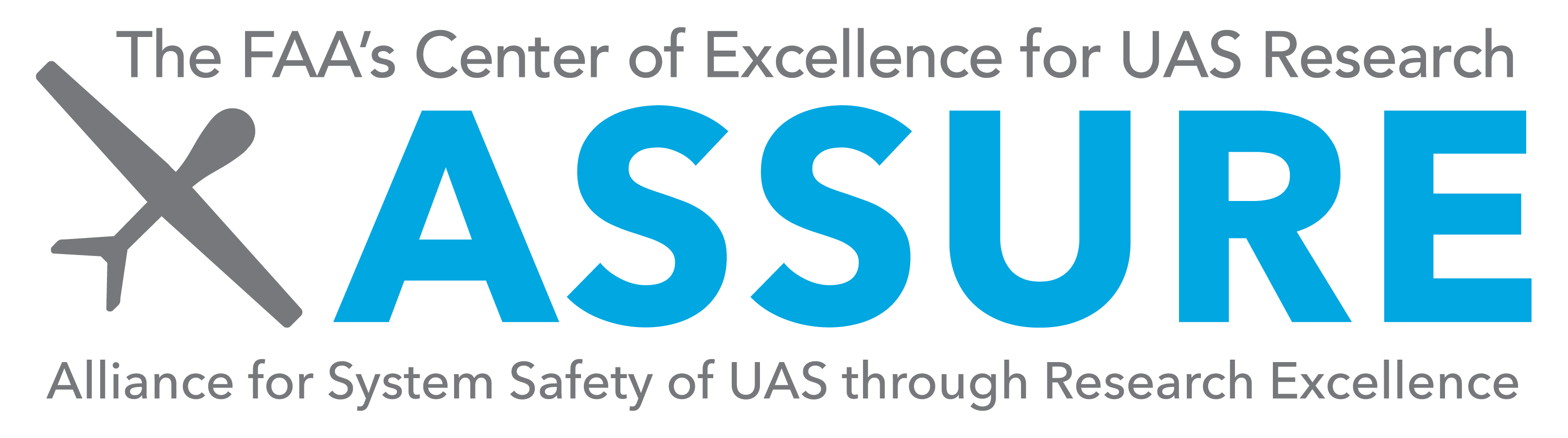}}}

\usepackage{amssymb}

\begin{titlepage}
\centering

\title{
\includegraphics[scale=0.45]{ASSURE-Logo.png}\\
\begin{figure}[h]
\centering
 {\includegraphics[width=.25\linewidth]{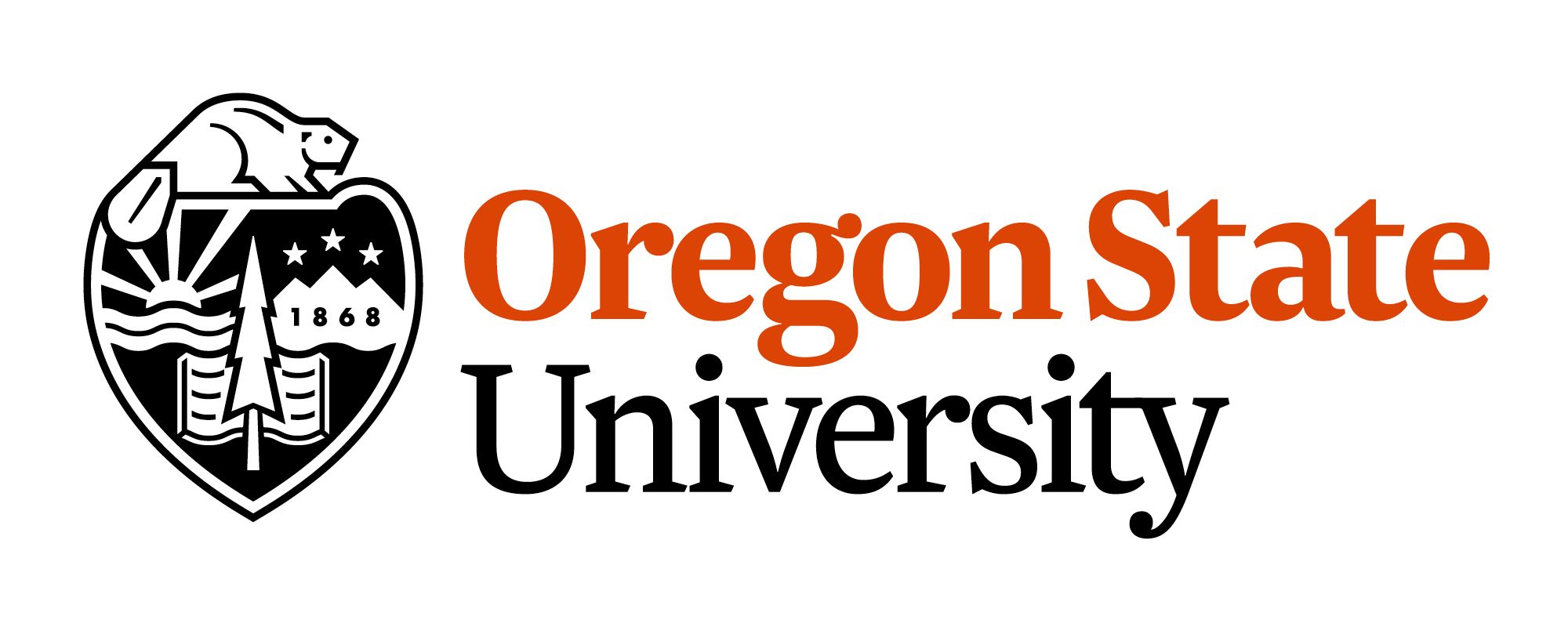}}
\end{figure}
\vspace{4em}
\Large\textbf{OSU-Wing PIC Phase I Evaluation: \\
Baseline Workload and Situation Awareness Results\\
\vspace{6em}
}}
\author{Julie A. Adams (julie.adams@oregonstate.edu),
Christopher A. Sanchez,\\
Vivek Mallampati,
Joshua Bhagat Smith,\\
Emily Burgess and
Andrew Dassonville}
\date{August 30, 2024}
\end{titlepage}

\begin{document}
\maketitle

\pagenumbering{roman}
\addtocounter{page}{1}
\clearpage

\setcounter{secnumdepth}{4}
\setcounter{tocdepth}{4}
\renewcommand{\contentsname}{Table of Contents}
\phantomsection
\addcontentsline{toc}{section}{Table of Contents}
\tableofcontents

\newpage
\renewcommand{\listfigurename}{Table of Figures}
\phantomsection
\addcontentsline{toc}{section}{Table of Figures}
\listoffigures

\newpage
\renewcommand{\listtablename}{Table of Tables}
\phantomsection
\addcontentsline{toc}{section}{Table of Tables}
\listoftables

\newpage
\renewcommand{\contentsname}{Table of Acronyms}
\phantomsection
\addcontentsline{toc}{section}{Table of Acronyms}

\textbf{\Large{Table of Acronyms}}

\begin{longtable}{ p{.17\textwidth} p{.72\textwidth}} 
\endfirsthead
 \textbf{Acronym} & \textbf{Meaning}  \\

\endhead 
 \textbf{Acronym} & \textbf{Meaning}  \\
  ADS-B & Automatic Dependent Surveillance - Broadcast \\
  ANOVA & Analysis of Variance \\
  AOI & Area of Interest \\ 
  DAA & Detect and Avoid\\
  DARPA & Defense Advanced Research Projects Agency\\
  FAA & Federal Aviation Administration  \\
  KDFW & Dallas-Fort Worth Airport identifier \\
  METAR & Meteorological Terminal Aviation Routine Weather Report \\
  OSU & Oregon State University \\
  PIC & Pilot In Command \\
  SA & Situation Awareness \\
  SA$_{2}$ & Situation Awareness Level 2 Probe \\
  SA$_{3-4:30}$ & Situation Awareness Level 3 Probe at minute 4:30\\
  SA$_{3-8:00}$ & Situation Awareness Level 3 Probe at minute 8\\
  std & Standard Deviation \\
  UAS & Unmanned Aircraft System  \\

\end{longtable}

\clearpage

\pagenumbering{arabic}
\setcounter{page}{1}
\phantomsection
\addcontentsline{toc}{section}{Executive Summary}
\section*{Executive Summary}

The common theory is that human pilot's performance degrades when responsible for an increased number of uncrewed aircraft systems (UAS). This theory was developed in the early 2010's for ground robots and not highly autonomous UAS. It has been shown that increasing autonomy can mitigate some performance impacts associated with increasing the number of UAS. Drone light shows are a compelling example and delivery drone systems, such as Wing's, have demonstrated reliable autonomy and the ability of pilot's to manage multiple UAS simultaneously. 

Overall, the Oregon State University-Wing collaboration seeks to leverage Wing's capabilities to understand what factors, other than the number of UAS, negatively impact a pilot's ability to maintain responsibility and control over an assigned set of active UAS. The collaboration identified a two phase user evaluation process to understand the factors that impact human performance and how those factors impact performance. The Phase I evaluation establishes baseline data focused on Wing's operating procedures at the time the evaluation was designed in late 2023 and early 2024, as well as when the number of UAS and the number of nests increase. This evaluation focuses on nominal operations as well as crewed aircraft encounters and adverse weather changes. 

The Phase I evaluation's 60 minute trials incorporated nominal conditions, a single crewed aircraft encounter, a condition in which two crewed aircraft encounters occurred one after the other and remained simultaneously for a period of time, and an adverse weather condition. The number of active UAS and the number of nests were manipulated. The evaluation assessed subjective and objective workload, subjective situation awareness, and provided objective metrics of the pilots' attentional focus and interface interactions. 

The hypothesis was that manipulating the number of active UAS, the number of UAS nests and the unexpected conditions will not impact operator performance, where performance is measured by workload, situation awareness and focus of attention. The results demonstrate that the pilots were actively engaged in the tasks and had good to very good situation awareness across all conditions.   Manipulation of the conditions did not result in any significant differences in overall workload, which generally remained in the normal range. The pilots visually focused the most on the ADS-B display, followed by the Wing user interface. The Weather display did receive increased attention during the adverse Weather task.  Pilots also interacted with the ADS-B display and Wing interface nearly exclusively, and in ways that demonstrated they were gathering pertinent information relevant to their job duties. These interactions were largely consistent across tasks and trials in nature and what PICs interacted with, although the rate of interactions was affected by unexpected events.   

These results debunk the traditional theory that increasing the number of UAS is detrimental to pilot's performance. The purpose of the Phase I evaluation was to generate these baseline results prior to conducting the Phase II evaluation. The Phase II evaluation will identify and manipulate factors that are expected to impact pilot performance when deploying multiple UAS, in the hopes of supporting the development of mitigation strategies that optimize safety and efficiency.  

\clearpage
\setcounter{table}{0} 
\renewcommand{\thetable}{\arabic{table}}
\section{Introduction} 
\label{sec:Introduction}
Many organizations believe that the human-to-robot (uncrewed aircraft) ratio is the key metric driving issues of scaling multiple uncrewed aircraft systems (UAS) for integration into the national air space. While this ratio can be a relevant characteristic that impacts human performance, from a human factors perspective it is unlikely to be the case given the dynamic and resilient nature of Wing’s delivery UAS.  Highly automated systems such as Wing's, include the ability to respond to unexpected events predictably, rapidly, and without the need for human intervention. It is therefore more likely that there are other relevant task, cognitive, and perceptual factors that better predict human performance when a pilot in command (PIC) serves as a supervisor, or some similar role.

Unfortunately, there is a lack of conclusive data sets and results that comprehensively analyze the breadth of potential factors that contribute to performance decrements within such autonomous UAS systems. Most existing results are the outcomes of much simplified highly constrained and controlled human subjects evaluations, which are less ecologically valid and predominantly produce only subjective or qualitative results that cannot definitively capture objective changes in workload, decision making, and other aspects of complex human performance \cite{BassetalHFES2022,BassetalA-26LitReview2021,GlavanetalHMS2022,AdamsetalA-26Final}. 
Thus, there is a need to collect robust, objective data that more appropriately represent human performance within actual real-world contexts and systems. This type of data collection will be useful for several reasons, not the least of which being their utility in a) informing regulators (i.e.,\ FAA) of the true scope of safe and reliable operations, and b) developing a standard for evaluating similar systems, such that scaling UAS deployments can be done safely in the national airspace. Oregon State University (OSU) and Wing have embarked on an effort to understand what human performance factors may impact a PIC's human performance when supervising delivery drones (i.e.,\ UAS) in a highly automated environment. 

OSU and Wing designed a two phase experiment. The first phase (Phase I) focused on gathering human factors data based on the number of active UAS, number of nests, and the impact of unexpected events for which a single PIC can be responsible. Phase I was designed to provide baseline data, given that these particular factors were predicted to have little if any impact on a PIC's performance. The Wing PIC delivery interface and Wing's UAS delivery simulator were used for the Phase I evaluation. This Phase I evaluation will inform the design of the second phase (Phase II) is intended to vary factors (e.g.,\ area size, geographical diversity in an area) that may impact a human’s performance when deploying multiple UASs. 

The use of Wing’s UAS delivery simulator to emulate the deployment of larger numbers of UAS in differing conditions enables the evaluation of mission deployments consistent with current and future industry aspirations, and facilitates collecting objective human performance results.  

\section{Experimental Design} \label{sec-ExprDesign}

The Phase I evaluation was designed to gather an objective workload and relevant dataset as a baseline for comparison to the eventual Phase II evaluation that will investigate a broader set of factors (e.g.,\ area size, density of UASs in an area, geographical diversity in an area). The \textit{general research question} for Phase I was straightforward: Is the PIC's performance impacted by the number of nests, number of UAS, or common unexpected events?  

\begin{figure}[htb]
\centering
\includegraphics[width=.75\linewidth, keepaspectratio]{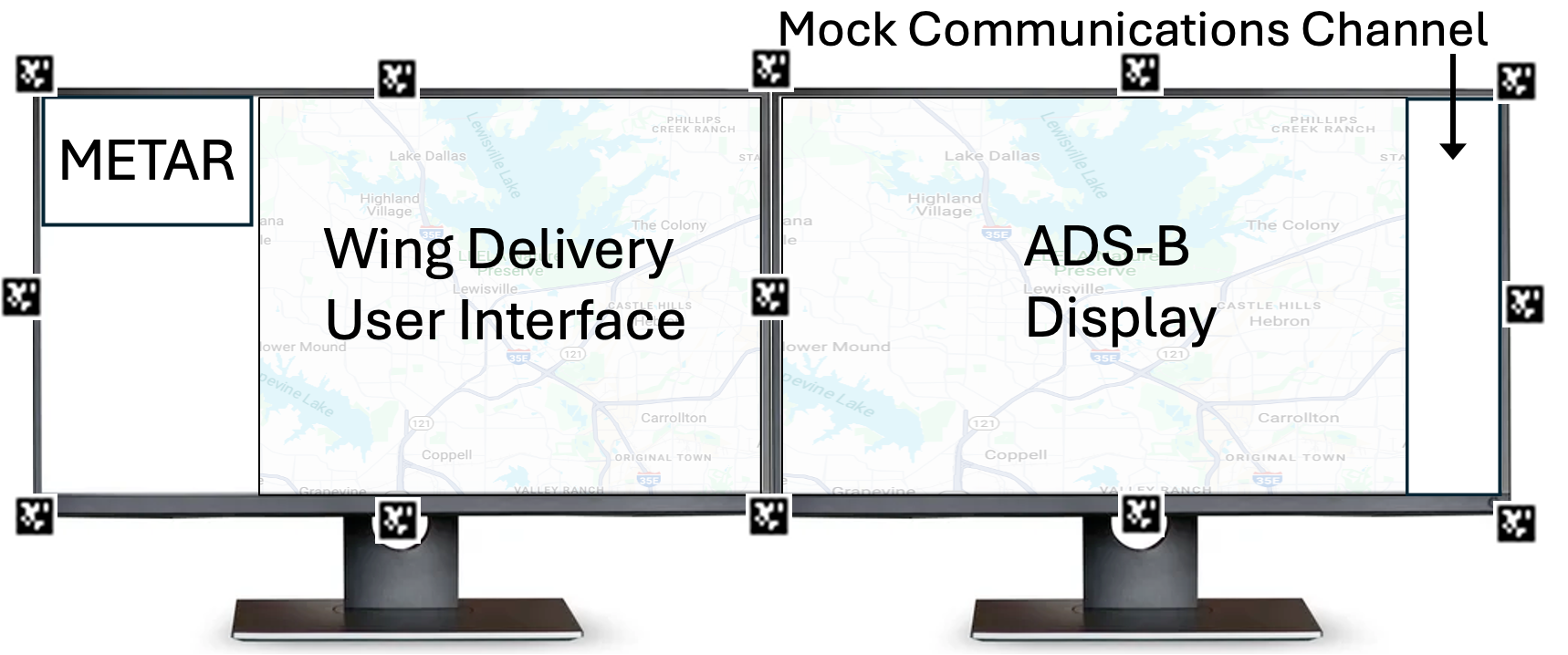}
\caption{The monitors with the respective open windows for each trial session. The eye tracking analysis AprilTags were aligned on the outsides and the center of the monitors.}
\label{fig:Monitors}
\end{figure}

The PICs completed tasks similar to their normal work duties. The tasks required simultaneously using two monitors positioned (shown in Figure \ref{fig:Monitors}) to supervise the active UAS while also monitoring the airspace, weather, and communications. The overall PIC responsibilities differed slightly from their current operational duties in that they did not actively communicate with personnel at the nests for any reason and the operational area in the Dallas-Fort Worth area was spatially larger (37 miles x 20 miles) with a higher number of nests and active UAS. 

Wing's UAS delivery simulator emulated the assignment and execution of deliveries by the UAS (i.e.,\ active UAS), and this information was provided to the PICs via Wing's existing delivery user interface. The standard weather tools used during normal flight operations provide real-time weather conditions that cannot be manipulated, and thus were not integrated into the current evaluation. Instead, the weather display presented a text-based representation of nominal operational weather conditions for the Dallas-Fort Worth (KDFW) airport's Meteorological Terminal Aviation Routine Weather Report (METAR) on a Google Slide. The Google Slide text was updated with the adverse weather conditions at a specified time. The left monitor displayed the Wing delivery user interface presenting the active UAS, with the METAR weather information in the top left, as shown in Figure \ref{fig:Monitors}. 
The Automatic Dependent Surveillance–Broadcast (ADS-B) tool's display consumed much of the right monitor's display, with a mock communications channel along the right side. The ADS-B provided real-time crewed aircraft traffic within the filtering criteria for the area of operation. Scripts were used to inject simulated crewed aircraft traffic encounters.

\subsection{Independent Variables}

\begin{figure}[htb]
\centering
\includegraphics[width=.5\linewidth, keepaspectratio]{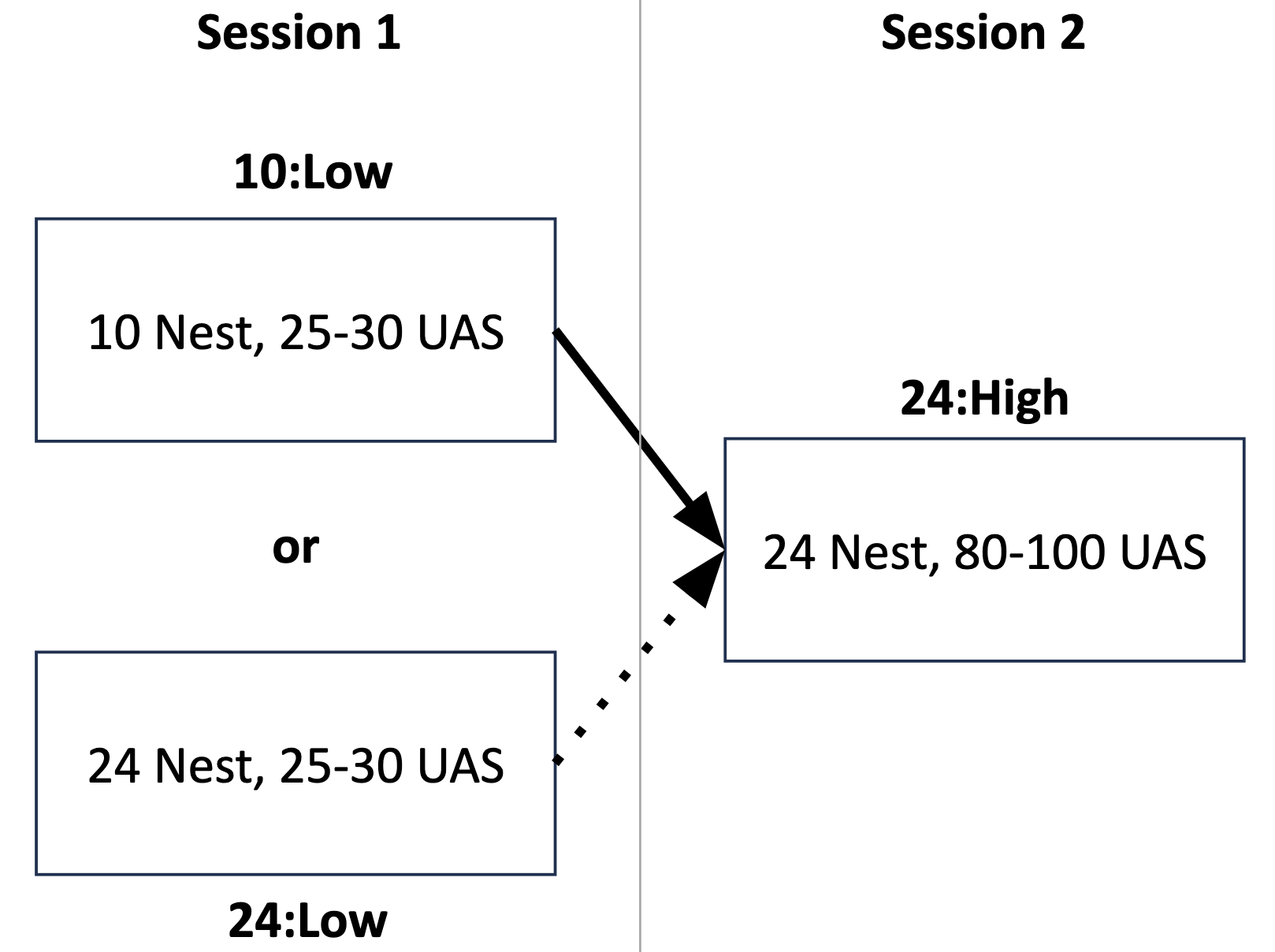}
\caption{An overview of the three trial conditions by session.}
\label{Fig:TrialOverview}
\end{figure}

Multiple independent variables were manipulated during the Phase I evaluation. The number of active nests was either 10 or 24, where some nests contained differing numbers of UAS. The size of the available fleet was dependent on the number of UAS in each active nest (i.e.,\ 10 Nests = 120 UAS, 24 Nests = 446 UAS). The PIC ratio, a Wing simulator parameter that modulates the number of active UAS (i.e.,\ in flight) for which a single PIC is responsible was set to provide a low number of UAS with a target range of 25-30 UAS in flight, and a high number of UAS with a target range of 80-100 UAS in flight. These variables resulted in three trial conditions: 10:Low, 24:Low and 24:High, as outlined in figure \ref{Fig:TrialOverview}. 

The 10:Low condition had 10 active nests and an allowable maximum PIC ratio of 45, which was expected to result in 25-30 active UAS (Low). The 24:Low condition increased the number of nests to 24 with a PIC ratio of 35 to regulate the number of active UAS to be between 25-30. The 24:High condition also had 24 nests, but instead set the PIC ratio to 200 to provide an expected number of 80-100 active UAS (High). 

Each PIC completed two trials, each of which was 60 minutes in length. The PICs were randomly assigned to either the 10:Low or 24:Low conditions for their first session, as shown in Figure \ref{Fig:TrialOverview}. All PICs completed the 24:High condition during the second session. 

\begin{figure}[htb]
\centering
\includegraphics[width=.99\linewidth, keepaspectratio]{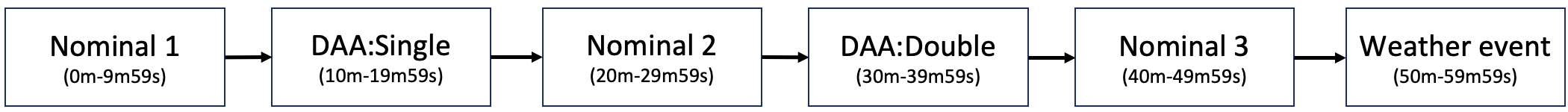}
\caption{The task order and timing for each trial within all tasks.}
\label{Fig:TaskOrder}
\end{figure}

Each 60-minute trial was composed of six, ten minute tasks that always occurred in the same order, as shown in Figure \ref{Fig:TaskOrder}. Three nominal task segments were interspersed between unexpected event tasks. The first nominal task included a ramp up period for the 10:Low and 24:Low conditions as UAS were deployed. The UAS delivery simulator was started approximately 30 minutes prior to the start of the 24:High condition to make the first nominal trial's ramp up period similar to the X:Low UAS conditions' ramp up. 

The unexpected event tasks represent common situations that the PICs encounter during a normal routine shift. These events included two detect and avoid (DAA) scenarios, and the onset of adverse weather conditions. The DAA:Single task involved a single crewed aircraft encounter that the PIC must monitor and potentially respond to. The intent was that the encounters were not to result in the PIC pausing Wing's operations. The DAA:Double event task had two such crewed aircraft encounters. These DAA tasks required a script be manually activated to modify the ADS-B display to include the specific crewed aircraft encounter. The PICs were responsible for detecting the crewed aircraft encounter on the ADS-B display, monitoring said aircraft relative to the active Wing UAS, and taking appropriate necessary actions. The DAA:Single encounter  was expected to occur approximately three minutes into the task and last through this task's duration. The first crewed aircraft encounter of the DAA:Double task was expected to occur at approximately two minutes into the task, with the second encounter expected to occur at approximately four minutes into the task. The DAA:Double task's two crewed aircraft encounters persisted on the ADS-B display for at least four minutes of the total task duration. The X:Low trials used the same crewed aircraft encounter script, as PICs only completed one of these trials. This script's crewed aircraft encounter differed from that used in the 24:High condition. These differences were intentional to prevent participants from simply recalling the nature of the previous crewed aircraft encounter, including their reaction to said event, which may produce an unrealistic estimation of their experienced workload.

\begin{figure}[!htbp]
    \centering
    \begin{subfigure}[t]{0.45\textwidth}
        \includegraphics[width=.99\textwidth]{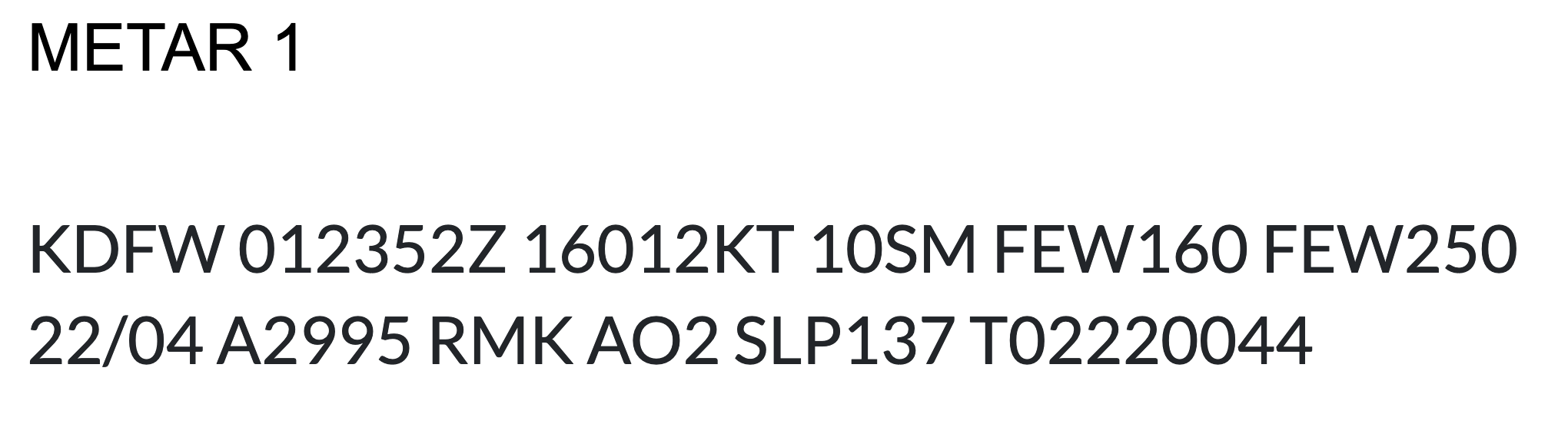}
        \caption{The nominal weather.}  
        \label{fig:NominalMETAR}
    \end{subfigure}
    \begin{subfigure}[t]{0.45\textwidth}  
        \includegraphics[width=.99\textwidth]{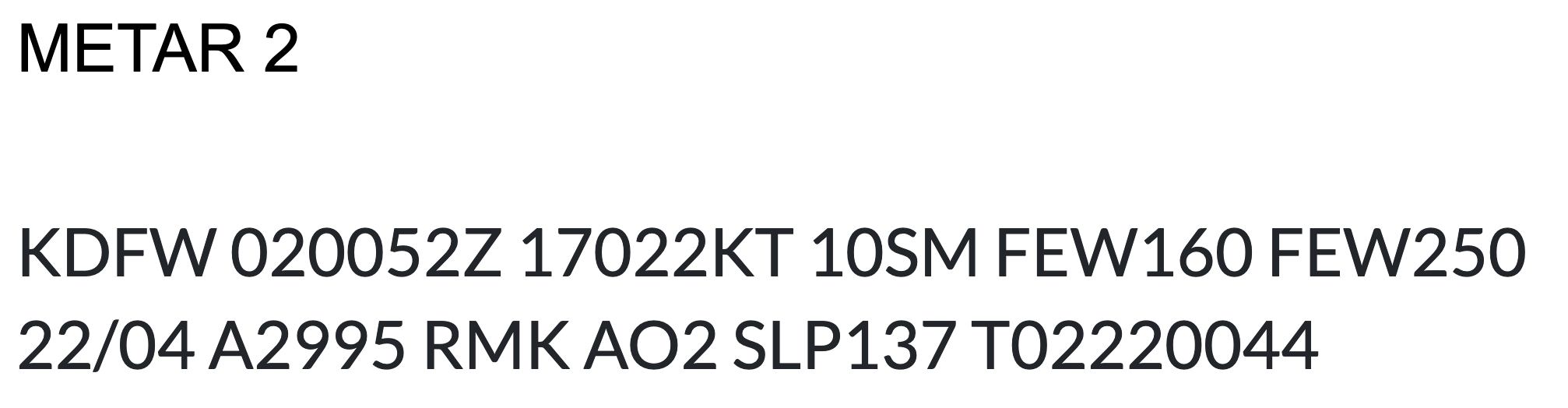}
        \caption{The high winds weather.}   
        \label{fig:HighWindsMETAR}
    \end{subfigure}
    \caption{The X:Low weather METARs.}
    \label{fig:METAR}
\end{figure}

The PICs were informed that the Weather METAR was to update at 45 minutes past the top of the hour ($\pm 10$\ minutes). An experimenter provided a new METAR that represented an adverse Weather condition at approximately 52 minutes into the trial. The X:Low trials' updated Weather represented high wind conditions, while the 24:High trials encountered low ceiling conditions. An example of the X:Low trial METARs are provided in Figure \ref{fig:METAR}. The PICs were expected to pause the system operations based on the adverse Weather update. 

\subsection{Dependent Variables}
The dependent variables included the physiological data from the wearable sensors, the Wing simulator log files, the eye tracker, screen capture videos and camcorder videos, as well as the subjective \textit{in situ} workload and situation awareness (SA) probes. The subjective workload and situational awareness probe responses were collected via Qualtrics, as was the demographic questionnaire. Qualtrics also facilitated capturing the time PICs required to answer each in situ workload rating and SA probes. The Wing simulator log files provided essential data regarding the number of active UAS at each time step, including the number of active UAS that were: taking off, en route to delivery, returning to the nest, or landing. 

The physiological data was used to estimate the cognitive, visual, speech, auditory, gross motor, fine motor, and tactile workload components, as well as overall workload levels.  Workload estimates between 20 and 59 are classified as normal workload, with values $\geq 60$ representing an overload state, and values $< 20$ representing an underload state. 

Wearable sensors' data streams corresponded to each respective workload component. The wearable sensors used for this evaluation are provided in Table \ref{tab:tr_metrics}. The Biopac Bioharness is worn around the chest, Pupil Labs Neon eye tracker are glasses, and the a Shure microphone is worn as a headset. The two Myo arm bands are worn on each forearm,and a Reed decibel meter sits on the table. The sensor streams were recorded and processed post-hoc using OSU's multi-dimensional workload estimation algorithm to estimate  overall workload and each workload components (e.g.,\ cognitive, tactile) values on a scale from zero to 100. The eye tracking data was also leveraged to understand PICs' locus of attention and engagement as well as the specific interactions with the available systems (e.g.,\ ADS-B). 

\begin{table}[!htbp]
\centering
\caption{The wearable sensors by metric and workload component.}
\label{tab:tr_metrics}
\resizebox{\textwidth}{!}{%
\begin{tabular}{|c|c|c|c|c|c|c|c|c|}
\hline
\textbf{Sensor} & \textbf{Metric} & \textbf{Cognitive} & \textbf{Speech} & \textbf{Auditory} & \textbf{Visual} & \textbf{\begin{tabular}[c]{@{}c@{}}Gross\\ Motor\end{tabular}} & \textbf{\begin{tabular}[c]{@{}c@{}}Fine\\ Motor\end{tabular}} & \textbf{Tactile} \\ \hline
 & Heart Rate &  &  &  &  & \cellcolor[HTML]{9B9B9B} &  &  \\ \cline{2-9} 
 & Heart Rate Variability & \cellcolor[HTML]{9B9B9B} &  &  &  &  &  &  \\ \cline{2-9} 
 & Respiration Rate &  &  &  &  & \cellcolor[HTML]{9B9B9B} &  &  \\ \cline{2-9} 
\multirow{-4}{*}{Bioharness} & Postural Magnitude &  &  &  &  & \cellcolor[HTML]{9B9B9B} &  &  \\ \hline
 & Pupil Diameter & \cellcolor[HTML]{9B9B9B} &  &  & \cellcolor[HTML]{9B9B9B} &  &  &  \\ \cline{2-9} 
 & Blink Latency & \cellcolor[HTML]{9B9B9B} &  &  & \cellcolor[HTML]{9B9B9B} &  &  &  \\ \cline{2-9} 
 & Blink Rate & \cellcolor[HTML]{9B9B9B} &  &  & \cellcolor[HTML]{9B9B9B} &  &  &  \\ \cline{2-9} 
 & Fixations &  &  &  & \cellcolor[HTML]{9B9B9B} &  &  &  \\ \cline{2-9} 
\multirow{-5}{*}{Neon Eye Tracker} & Saccades &  &  &  & \cellcolor[HTML]{9B9B9B} &  &  &  \\ \hline
 & Voice Intensity &  & \cellcolor[HTML]{9B9B9B} &  &  &  &  &  \\ \cline{2-9} 
 & Voice Pitch &  & \cellcolor[HTML]{9B9B9B} &  &  &  &  &  \\ \cline{2-9} 
 & Speech rate &  & \cellcolor[HTML]{9B9B9B} &  &  &  &  &  \\ \cline{2-9} 
 & MFCCs &  & \cellcolor[HTML]{9B9B9B} & \cellcolor[HTML]{9B9B9B} &  &  &  &  \\ \cline{2-9} 
\multirow{-5}{*}{Microphone} & Spectrogram &  & \cellcolor[HTML]{9B9B9B} & \cellcolor[HTML]{9B9B9B} &  &  &  &  \\ \hline
\begin{tabular}[c]{@{}c@{}}Reed\\ Decibel Meter\end{tabular} & Noise level &  &  & \cellcolor[HTML]{9B9B9B} &  &  &  &  \\ \hline
Myo armband & sEMG &  &  &  &  &  & \cellcolor[HTML]{9B9B9B} & \cellcolor[HTML]{9B9B9B} \\ \hline
\end{tabular}
}
\end{table}

The subjective \textit{in situ} workload probes were administered during each task at 3 and 6 minutes. The PICs rated their perceived workload for each workload component on a scale from 1 (exceptionally low) to 7 (exceptionally high). 

In situ situation awareness (SA) probes were administered during each task. The SA probes represented  SA Levels 2 and 3 (comprehension and prediction, respectively) and no SA Level 1 probes (perception) were asked. The SA Level 2 probe was asked at the 2 minute mark, while two different SA Level 3 probes were asked at the 4:30 minute and 8 minute marks. The SA level 2 (SA$_{2}$) probe asked: ``What aircraft [provide tail number] in the outer ring do you consider the most erratic?'' The SA level 3 probe at the 4:30 minute mark (SA$_{3-4:30}$ was ``Do you expect to change the system status in the next three minutes?,'' while the probe at 8 minutes (SA$_{3-8:00}$) was ``Do you expect the system operability to change in the next two minutes?.'' The three probes were the same across all tasks and all trials. The responses to all SA probes were analyzed to determine the response accuracy.

During the Nominal and Weather tasks, a correct SA$_{2}$ probe response had a valid explanation (i.e.,\ identifying any aircraft present on the ADS-B display) given that there were no crewed aircraft that were potential encounters. A correct response during the DAA tasks was the crewed encounter aircraft's identifier, which required determining if the encountered crewed aircraft(s) was present on the ADS-B display when the SA$_{2}$ probe was asked. The ADS-B display presented crewed aircraft that were within the area of operation, and was updated in real-time. A script was used to inject the specific crewed aircraft encounter(s) onto the ADS-B display. The script was activated by Wing's experimental personnel, and there was some temporal variation in when the script activated. ADS-B display capture videos were coded to identify the exact time at which each crewed aircraft encounter appeared for a given PIC and trial. If an encounter was present during a SA$_{2}$ probe, the correct response was the crewed vehicle's identifier. If no crewed encounter was in fact present, then providing an identifier for any crewed aircraft that was present on the ADS-B was coded as correct. 

The SA Level 3 probes required a ``Yes'' or ``No'' response. The correct response in most cases was ``No'', as the system status (SA$_{3-4:30}$) or system operability (SA$_{3-8:00}$) was not expected to change. The DAA:X tasks did not require pausing the system; however, the Weather task was expected to result in the PIC pausing the system prior to the first SA$_{3}$ probe, as the weather was updated at approximately 2 minutes into the Weather task. 
A few PICs during the Weather trials did not pause the system by the time of \textit{either} SA$_{3}$ probe, which is likely a result of the PIC not noticing the change in the METAR information. Only if the PIC had paused the system and answered ``No'' was the response coded as correct. If the system was not paused, and the PIC answered ``No'', the answer was coded as incorrect. Additionally, any other response (i.e., ``maybe'') was also coded as incorrect. 

\begin{figure}[htb]
\centering
\includegraphics[width=.5\linewidth, keepaspectratio]{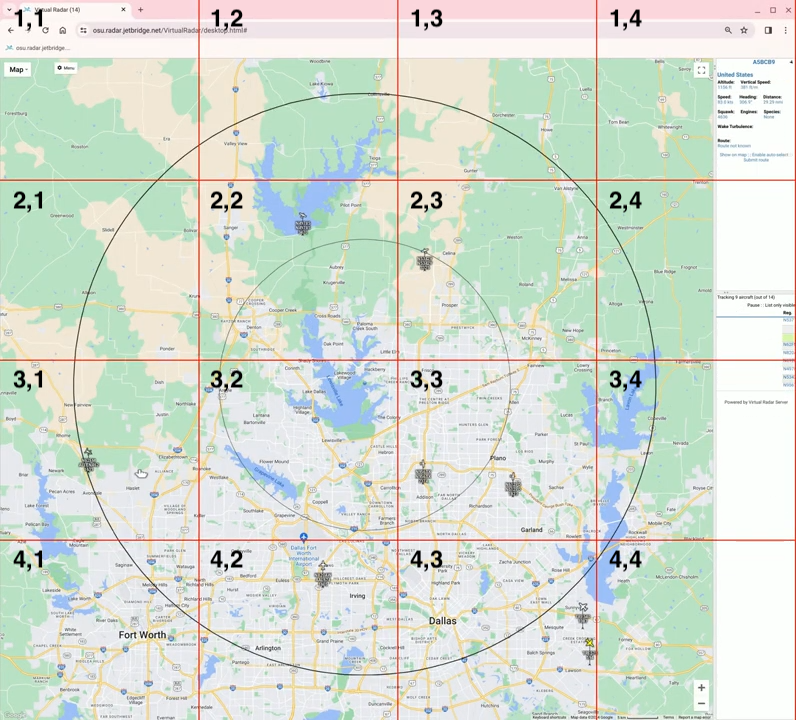}
\caption{The ADS-B display with the AOI numbered $(X,Y)$ cells of the 4 x 4 grid.}
\label{fig:GridCells}
\end{figure}

Additional eye tracking analyses were also conducted that examined fixation count and fixation duration for specific spatial locations on the displays, referred to as Areas of Interest (AOI). These AOIs were defined using thirteen AprilTags from the tag36h11 family. The 2 x 2 inch AprilTags were placed on each monitor, as shown in Figure \ref{fig:Monitors}. Each AprilTag was placed on the monitor's edge, and did not obstruct any portion of the screen. Broad AOI mapping resulted in five areas: the Weather display, the Wing delivery interface display, the ADS-B display, the chat display, and other (e.g., fixation on any area of the screen not specified within the other AOIs, or fixations outside the monitors). Additional more fine-grained AOIs were specified \textit{within} the ADS-B display, which was overlaid with a 4 x 4 grid to provide more exact consideration of where PICs were looking within the ADS-B display. The cell identifiers defined by this grid overlay on the ADS-B are provided in Figure \ref{fig:GridCells}. 

Eye tracker camera and camcorder videos, along with the ADS-B screen captures were analyzed to identify user interactions with the various displays and resources.  The five displays and resources were: ADS-B, Wing Interface, METAR, Chat, and Offscreen materials.  The Offscreen materials were presented digitally, and were presented either via paper documentation or verbally.  Interactions with the five computerized displays were executed with a computer mouse, and were broadly coded for the corresponding activity. The ADS-B contained five broad actions: (1) obtaining crewed aircraft information by either mouse click or mouse hovering, (2) using the mouse arrow cursor as an external reference point or indicator, (3) adjusting the ADS-B menu settings, (4) zooming the ADS-B display in or out, and (5) panning the ADS-B view up/down/left/right.  The Wing Interface likewise contained five broad actions: (1) obtaining Wing UAS information by clicking or scrolling through the aircraft menu, (2) using the mouse arrow cursor as an external reference point or indicator, (3) zooming the interface in or out, (4) adjusting settings for the Wing interface, and (5) panning the Wing interface view up/down/left right.  Interactions with Offscreen materials were coded for either: (1) Reading Paper Documentation, or (2) Asking a verbal question.  The Chat was coded for whether PICs selected or clicked a message.  The METAR was coded for any highlighting of information.  

\subsection{Procedure}
Each session occurred in a Wing pilot training facility.  Wing's PICs provided informed consent upon arriving for their first session, after which PICs completed a computerized demographic questionnaire that captured their age, gender, education, technology use, and pilot experience. A short working memory capacity (Symmetry Span; \cite{oswald2015development}) test and a test of their visuospatial processing, specifically visuospatial visualization (Paper Folding Task, Part 2; \cite{educational1963kit}) followed. The Symmetry Span test evaluates how well PICs can monitor and control their attention, while the Paper Folding task measures how well PICs can manipulate spatial information mentally.  After completing these initial assessments, PICs were fitted with the wearable sensors and sensors the required calibration (e.g.,\ Myos) were calibrated. The PICs were asked to review the adapted PIC Reference Sheet and Procedures Manual (provided by Wing personnel) that provided the flight specifications for the mock shift. Once this review was completed and any questions were answered, the PICs began trials.

Half of the participants completed one of the X:Low conditions during their first session, and all completed the 24:High condition for their second session. Both sessions were performed on the same day with at least 1.5 hours between sessions. Each session, including setup and the sensor donning and calibration, took approximately 90 minutes in total. After completing the final session, PICs were debriefed regarding the nature of the study. 

\subsection{Participants}
Wing trained and Part 107 certified PICs were the participants. A total of seven participants completed the evaluation, but one participant's results were removed due to incorrect scenario system settings. An \textit{a-priori} power analysis using G*Power with a medium effect size ($f$ = 0.25), $\beta = 0.20$, and $\alpha = 0.05$, and a modest correlation between factors ($r = 0.50$), revealed that detecting within-between interaction effects with six PICs has a power of 0.89.

All six PICs have normal vision and passed the Ishihara color blindness test as eligibility for their Wing position. The six PICs had a mean age of 35.67 years (range: 28-45; all males), with three holding a Bachelor's degree, two holding a Master's degree, and one having some college experience. The average amount of time spent in their current position at Wing was 1.5 years (range: one month - 4.5 years, at the time of the evaluation), while the average time spent supervising UAS was 3.23 years (standard deviation: std = 2.84). Related to technology use, PICs spent an average 40.15 hours (std = 13.57) on the computer weekly and 1.67 hours (std = 2.07) playing video games. Three PICs held a pilot's license with an average of 235 flight hours. The average working memory score (maximum score of 24) for the six PICs was 16.33 (std = 4.97), and the average visuospatial processing score (maximum score of 10) was 4.33 (std = 1.86), both of which are well within the normal range.

\subsection{IMPRINT Pro Model}

IMPRINT Pro (Improved Performance Research Integration Tool) is a cognitive modeling tool that supports manpower, personnel and human systems integration. IMPRINT Pro can handle dynamic, stochastic, discrete events. The tool models interactions between humans and systems, informs system requirements, identifies human performance-driven system design constraints, and evaluates personnel training capabilities and manpower requirements under environmental stressors. IMPRINT Pro permits constructing complex task networks, with nodes organized sequentially, concurrently, and hierarchically. IMPRINT Pro plugins provide additional capabilities (e.g.,  unmanned systems, fatigue, and training effects).

\subsubsection{Model Development}
\label{subsubsecModelDev}
The IMPRINT Pro tool was developed for purposes different from supervising multiple UAS and uses a linear model of overall workload. This linear model results in the same workload being added for each new UAS the human is assigned, irrespective of the mission domain and UAS capabilities. However, based on field work \cite{Adamsetal2023JFR} this linear overall workload model is not representative of the expected actual workload for Wing's delivery drone use case. Insights from results developed for the FAA's ASSURE Center of Excellence project A-26 \cite{AdamsetalA-26Final} were leveraged to derive a relevant workload model.

UAS with reliable high autonomy levels, such as Wing's delivery UAS, transform the human's task from managing and interacting with each individual UAS to one focused on visually monitoring the UASs' actions and the airspace. IMPRINT Pro's limitations related to modeling and assessing human performance for such systems are not unique when the number of UASs increases. Workload ($w$) in scenarios involving the supervision of multiple UAS can be modeled using a logarithmic function, reflecting the efficiency of human visual scanning across increasing numbers of UAS. This approach acknowledges that as the number of UAS ($n$) increases, the workload grows at a rate less than linear due to the nature of visual search efficiencies. Assuming that workload varies linearly in relation to visual search time, a logarithmic function may be appropriate for modeling workload given:
\begin{equation}\label{eq1}
w = a + b \ln n,
\end{equation}
where $w$ is associated with the workload from monitoring a single UAS, and $b$ is the rate workload grows as additional UAS are incorporated. The workload for a single UAS, $a$, can be estimated using IMPRINT Pro’s existing workload rubrics. The rate parameter, $b$, needs to be estimated using another method, such as rescaling a logarithmic visual search time function of set-size, which can be achieved by factoring Equation \ref{eq1}:

\begin{equation}\label{eq2}
w = a (1 + \frac{b}{a} \ln n),
\end{equation}
and substituting a new parameter $r$, the logarithmic rate, for the quantity $\frac{b}{a}$ resulting in:
\begin{equation}\label{eq3}
w = a ( 1 + r \ln n).
\end{equation}

The difference between Equations \ref{eq1} and \ref{eq3} is that $b$ in Equation \ref{eq1} has dimensions $[workload items-1]$, whereas $r$ in Equation \ref{eq3} has dimensions $[items-1]$. This difference allows estimating the logarithmic rate directly from set-size gradients measured in units other than workload (e.g., search time). However, it is necessary to fit $r$ to the use case.

\begin{figure}[hbt]
\centering
\includegraphics[width=1\linewidth, keepaspectratio]{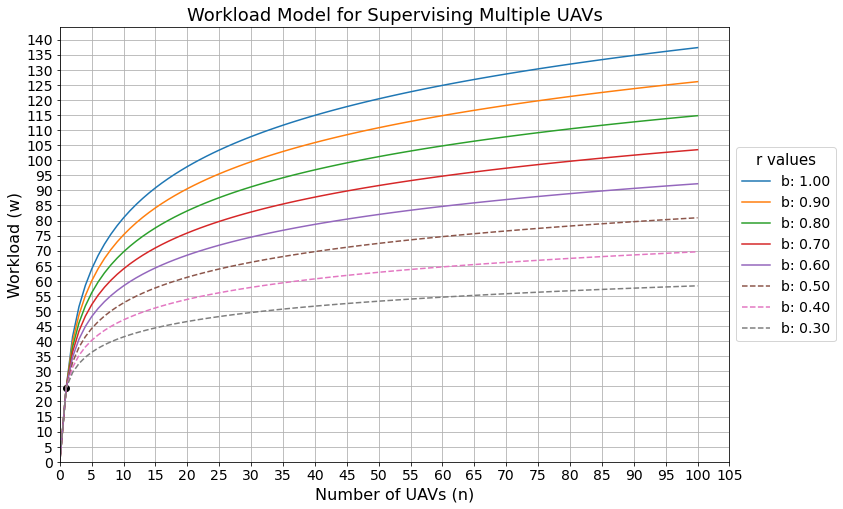}
\caption{The $r$ value analysis by the number of UAS for the nominal delivery use case.}
\label{fig:LogChart}
\end{figure}

It is necessary to define a more appropriate workload model than the one provided by IMPRINT Pro, as most alternative workload models do not appear to accurately represent workload for 1:N UAS scenarios \cite{Adamsetal2023JFR,AdamsetalA-26Final}. The approach for the developed IMPRINT Pro model, as represented in Equation \ref{eq3} requires choosing an appropriate log rate. Various log rates were analyzed for the nominal use case, as shown in Figure \ref{fig:LogChart}. The maximum expected target range was 80-100 active UAS for the 24:High trial. Based on Adams’ prior efforts with the DARPA OFFSET program \cite{Adamsetal2023JFR} and the FAA's ASSURE Center of Excellence project A-26 \cite{AdamsetalA-26Final}, the logarithmic rate ($r$) for the IMPRINT Pro workload model was set to 0.5.

\begin{figure}[!hbt]
\centering
\includegraphics[width=1\linewidth, keepaspectratio]{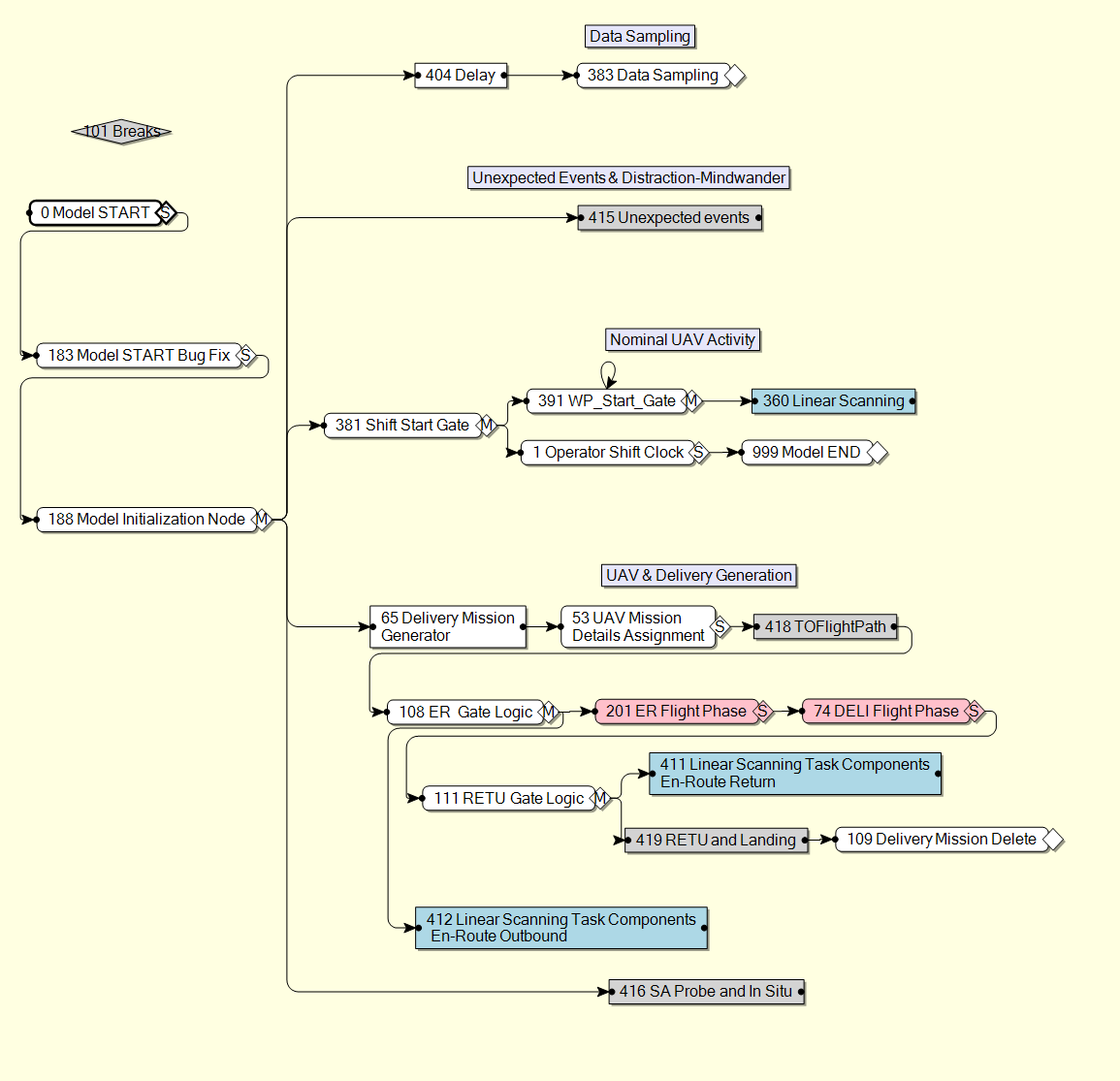}
\caption{The high-level developed Wing delivery use case IMPRINT Pro model.}
\label{fig:ImprintOverall}
\end{figure}

The workload model consists of three main elements (i.e., the nominal use case, the DAA crewed aircraft encounter use cases and the adverse Weather use case) along with the additional surveys. IMPRINT Pro models consist of various nodes that are parameterized to encode the use cases. The high-level overall model is presented in Figure \ref{fig:ImprintOverall}, with additional details hidden.

The nominal use case models the workload based on the PIC's normal activities without any unexpected events. The workload from monitoring multiple UAS follows the logarithmic function. This modeling allows for a more realistic representation of how workload increases with the number of UAS when considering the efficiency of human visual scanning. The IMPRINT Pro model must also consider unexpected events, which incorporates the additional workload due to the crewed aircraft encounters and adverse weather conditions. These scenarios require the PIC to use more visual and cognitive resources to successfully complete the task, increasing the overall workload. This aspect of the model acknowledges the increased demands on the PICs' attention and cognitive processing.

The IMPRINT Pro model also incorporates the human performance elements related to PICs responding to the in situ SA and workload rating probes during all tasks. The modeling of the probe responses is essential for accurately estimating the PIC's workload and ensuring the model reflects the real-world complexities of supervising multiple UAS.

The holistic workload model is intended to provide a more accurate estimation of a PIC's workload when supervising multiple UAS. This overall approach seeks to address the limitations of linear workload models and acknowledges the complexities and efficiencies of human visual scanning and cognitive processing in dynamic and challenging scenarios. 

The model results were generated for each of the trial conditions (10:Low, 24:Low and 24:High). Each model trial condition was customized to it's respective mean number of active UAS during the experimental data collection; as a result the 24:High condition was modeled to represent the differences in the number of active UAS for weeks 1 and 2. The model results were generated for a total of three runs for the 10:Low and 24:Low conditions and six runs for the 24:High condition (i.e.,\ three for each week).   

\subsubsection{Model Predictions}
\label{subsubsecModelResults}

The model predicted overall workload and each workload component (i.e., cognitive, visual, speech, auditory, gross motor, fine motor, and tactile).
The model predictions for 10:Low and 24:Low were the same; thus, these predictions were combined (i.e.,\ X:Low). The overall mean workload predicted by the model for the X:Low condition was 77.80 (std = 19.70), with the 24:High condition having a higher mean predicted workload of 91.83 (std = 23.32). 
All conditions predicted an overall workload $> 60$ (i.e.,\ overload). 


The number of active UAS during the 24:High trials varied by week. Week 2 had substantially fewer active UASs than week 1, but both have substantially more active UASs compared to the X:Low trials. No difference in the mean overall workload predictions existed between the two weeks of 24:High trials, as shown in Table \ref{Table:IMPRINTOverallTrial2}. The modeling assumptions anticipated higher workload level predictions for the 24:High trials due to the increased visual scan workload required when linearly scanning the larger number of active UASs.

\begin{table}[hbt]
\centering
\caption{The 24:High workload IMPRINT Pro prediction descriptive statistics [mean (std)]. } 
\label{Table:IMPRINTOverallTrial2}
\begin{tabular}{|c|c|c|}
\hline
\textbf{Overall} & \textbf{Week 1} & \textbf{Week 2} \\ \hline
91.83 (23.32) & 90.25 (22.40) & 93.41 (24.23) \\ \hline
\end{tabular}
\end{table}

The overall workload predictions for the three nominal tasks were generally identical, reflecting the stable demands of routine operations, as shown by trial and task in Table \ref{Table:IMPRINTWorkloadTasks}. The predicted overall workload noticeably increased with the unexpected DAA encounters (DAA:Single and DAA:Double) and the adverse Weather tasks. 
The double DAA encounters exhibit a substantially higher predicted overall workload, because the IMPRINT Pro model simulates the PIC managing the two DAA events independently. This independent handling of multiple crewed aircraft encounters introduces a compounding factor, resulting in the substantial difference. These predictions underscore the potential heightened complexity and demands a PIC may face when managing dynamic and challenging operational environments.

\begin{table}[hbt]
\caption{The overall workload IMPRINT Pro prediction descriptive statistics by trial and task. } 
\label{Table:IMPRINTWorkloadTasks}
\resizebox{\columnwidth}{!}{
\begin{tabular}{|l|c|c|c|c|c|c|}
\hline
\multirow{2}{*}{\textbf{Trial}} & \textbf{Nominal} & \textbf{DAA:} & \textbf{Nominal} & \textbf{DAA:}  & \textbf{Nominal} & \multirow{2}{*}{\textbf{Weather}} \\ 
 & \textbf{\#1} & \textbf{Single} & \textbf{\#2} & \textbf{Double}  & \textbf{\#3} & \\ \hline
X:Low  & 65.38 (8.08) & 82.20 (12.15) & 66.89 (6.53) &  101.62 (20.89) & 67.76 (7.37) & 84.96 (21.73)   \\ \hline
24:High & 81.99 (7.48) & 101.74 (11.33) &  82.56 (7.20) & 118.67 (18.07) &  79.79 (3.91) & 98.47 (37.16)  \\ \hline
\end{tabular}
}
\end{table}

The cognitive and visual components' IMPRINT pro predicted workloads, shown in Table \ref{Table:IMPRINTWorkloadComponents}, were the highest (i.e.,\ $> 30$). This outcome was expected, as the tasks have high cognitive and visual demands relative to the other workload components. The IMPRINT Pro model incorporated the auditory and verbal responses for the evaluation's in situ SA and workload probes, which are the primary drivers of the predicted auditory and speech workloads. Since the PICs are generally expected to be sitting or standing at a workstation during the trials, the IMPRINT Pro model predicted gross motor workload to be zero. The fine motor and tactile workload predictions result from mouse movements, and pressing mouse or keyboard buttons, respectively. These activities were modelled to occur more frequently during the unexpected events, when the PIC may pause future deliveries as necessary. 

\begin{table}[hbt]
\caption{The descriptive statistics for each components' IMPRINT Pro predicted workload by trial. The Gross Motor workload values were zero for all trials.} 
\label{Table:IMPRINTWorkloadComponents}
\resizebox{\columnwidth}{!}{
\begin{tabular}{|l|c|c|c|c|c|c|}
\hline
\multirow{2}{*}{\textbf{Trial}} & \multirow{2}{*}{\textbf{Cognitive}} & \multirow{2}{*}{\textbf{Visual}} & \multirow{2}{*}{\textbf{Speech}} & \multirow{2}{*}{\textbf{Auditory}}  & \textbf{Fine} & \multirow{2}{*}{\textbf{Tactile}} \\ 
 & & & & & \textbf{Motor} &\\ \hline
X:Low  & 33.44 (4.96) & 33.44 (4.96)  & 1.12 (1.53)  & 3.93 (3.61) &   8.22 (3.44) & 1.07 (1.35)    \\ \hline
24:High & 36.39 (7.02) & 40.22 (6.25) & 1.20 (1.66) & 4.21(3.81)   & 9.57 (3.57) & 1.12 (1.32) \\ \hline
\end{tabular}}
\end{table}


\section{ Results} \label{sec-Results}

The results are organized based on the three research questions. The number of active UAS during the trials are presented to provide appropriate context. Descriptive statistics (i.e.,\ mean and standard deviation) are provided for all results. Unless otherwise stated, Analyses of Variance (ANOVA) were conducted for the workload analysis and included within factors of: minute of trial, second, and block. Based on the ANOVA analysis, the \# of UASs (low or high) or \# of nests (10 or 24) were also included. These analyses were conducted for both Nominal tasks and for all unexpected events. 

The Wing UAS delivery simulator parameters were not set at the correct values for the one PIC, which resulted in the PIC pausing the system for a majority of the session. Three PICs' results are reported for the 10:Low trial, and the other three PICs completed the 24:Low trial. All six PICs' results are reported for the 24:High condition. 

The unexpected DAA encounter events were designed to have a consistent start time, but the activation of the associated script introduced variability. During the DAA:Single crewed aircraft encounter task across all trials, the event began on average at 148 seconds 
(s) (std = 54s), or 2 minutes (m) 28s into the 10m task. During the DAA:Double task, the first encounter occurred on average at 27s (std = 50s) 
into the task, with the second encounter occurring on average at 288s (std = 151s), or approximately 4m and 48s into the 10m task. The task specific mean timings are provided in Table \ref{Table:DAATiming}. 

\begin{table}[hbt]
\centering
\caption{The unexpected event times descriptive statistics, mean (std), represented by the number of seconds (s) into the 10 minute task by trial. }
\label{Table:DAATiming}
\begin{tabular}{|l|l|l|l|l|}
\hline
\multirow{2}{*}{\textbf{Trial}} & \multirow{2}{*}{\textbf{DAA:Single}} & \multicolumn{2}{|c|}{\textbf{DAA:Double}} & \multirow{2}{*}{\textbf{Weather}} \\ 
 & & \textbf{Encounter 1} & \textbf{Encounter 2} &\\ \hline
10:Low & 133 (8)    & 44 (14) & 345 (13) & 130 (16)\\ \hline
24:Low & 101 (87)    & 11 (77) & 117 (282) & 123 (3)\\ \hline 
24:High & 176 (39) & 23 (49) & 329 (61) & 140 (17) \\\hline 
\end{tabular}
\end{table}

There was less variability in the timing of the Weather events, as shown in Table \ref{Table:DAATiming}. The Weather update occurred on average at 133s (std = 16s) into the weather task, or at 2m and 13s. However, it did often take some time for the PICs to notice the Weather update, despite it being updated in the system view. 

\subsection{Number of Trial UASs}

Understanding the subsequent results requires investigating the number of UAS the PICs encountered in each trial. The PIC ratio was adjusted based on the trial condition and number of nests to achieve approximately the desired number of active UAS. The 10:Low trial incorporated 10 nests, a UAS fleet of 120, and a PIC ratio of 45, while the 24:Low trial increased the number of nests to 24, the total UAS fleet size of 446, and used a PIC ratio of 35. The desired outcome was to have the number of active UAS during these trials to be between 25 and 30. The number of active UAS for these two trials was generally as expected, as shown by the number of active UAS in Table \ref{Table:SimulatedNumUAS}. The minimum and maximum number of UAS were calculated from the time point where the first non-zero number of active UAS occurs after the start trial until minute 52, when the adverse Weather conditions were updated. The PICs did not necessarily notice the adverse Weather conditions at the time of the change. As such, there was variability in when they took the expected action of pausing the system and no additional UAS become active. Thus, the time of the Weather update was used as a consistent time point for this analysis.

\begin{table}[hbt]
\centering
\caption{The number of active UAS descriptive statistics by trial from the start of the trial until the Weather update at minute 52.} 
\label{Table:SimulatedNumUAS}
\begin{tabular}{|l|l|l|l|}
\hline
\multirow{2}{*}{\textbf{Trial}} & \multicolumn{3}{|c|}{\textbf{Active UAS}}  \\ 
& Mean (std) & Minimum & Maximum \\ \hline
10:Low & 27.33 (6.65)     & 0 & 33 \\ \hline
24:Low & 30.67 (1.25)     & 6 & 32  \\ \hline 
24:High & 76.33 (14.53)& 21 & 96 \\ \hline 
\end{tabular}
\end{table}

The 24:High trial's simulator parameters (i.e.,\ \# nests: 24, \# fleet UAS: 446, and PIC ratio: 200) were expected to generate between 80 and 100 active UAS, assuming the simulation was started 30 minutes prior to commencing the trial so that the simulation was able to ramp up to the desired number of active UAS. The overall number of active UAS during the 24:High trial was slightly lower than the minimum of 80, but with a high standard deviation. 
All data was collected over a two week period, and while this did not impact the X:Low trials, there were stark differences in the number of active UAS during the 24:High trial based upon week. The first week's data collection (4 PICs) resulted in a mean of 90.33 (std = 4.92, minimum = 25, maximum = 96) active UAS, while the second week (3 PICs) was substantially lower with a mean of 62.33 (std = 2.49, minimum = 21, maximum = 65) active UAS. A review of the experimental procedures and parameters was unable to reveal why such stark differences existed, and why the week two active UAS were substantially lower than the desired target number.  This said, there was still at least of doubling of UAS during the 24:High trials as compared to the X:Low trials, which provides a valid comparison point to examine how the number of UAS may impact workload.

\subsection{Locus of Attention}
\label{Sec:Focus}

The overall workstation incorporated the four interface displays, with the Weather and Wing Delivery user interface on the left monitor and the ADS-B and chat displays on the right monitor (see Figure \ref{fig:Monitors}). The experiment specific Reference Sheet and Procedures Manual indicated the PICs were to pause the delivery system if a crewed aircraft was below an altitude threshold and within (or on track) to enter the ADS-B display's inner circle. Under normal operational procedures, PICs are free to zoom their displays as preferred, which was retained during the evaluation. Wing's workstation hardware did not support video capture of both monitors' screens, thus, only the ADS-B display was recorded. However, the eye tracker's outward facing camera captured where a PIC was looking during the trials. 

The reported fixation durations are shorter than the full 60 minute session. Generally, the Pupil Labs Neon eye tracker's limitations include the loss of fixation data during rapid head movements (e.g.,\ shaking or turning) and PIC's physical eye tracker adjustments that partially cover the camera. The eye tracker also overheated twice and an experimenter accidentally unplugged the connection between the eye tracker and the recording device once. These three situations occurred during one of each condition for less than 90 seconds. 

Both X:Low trials each had fewer total fixations than the 24:High trials, as indicated in Table \ref{Table:FixCountOverall}, due to only three PICs completing each X:Low trial. However, the combined X:Low trials had 32,078 total fixations (i.e., 7.4\% more than the 24:High trials).  The total duration of 05:30:12  (i.e., hh:mm:ss) for the X:Low trials was 1.4\% higher than during the 24:High trials. The higher total X:Low trial fixations are likely due to PICs being uncertain what tasks they faced during the trials, as many PICs commented that they were unsure what to expect during their first session. 

\begin{table}[!hbt]
\centering
\caption{The total fixation counts and percentages by trial and display. } 
\label{Table:FixCountOverall}
\resizebox{\columnwidth}{!}{
\begin{tabular}{|c|c|c|c|c|c|c|c|c|c|c|c|}
\hline
 \multirow{2}{*}{\textbf{Trial}} & \multicolumn{2}{c|}{\multirow{2}{*}{\textbf{METAR}}} & \multicolumn{2}{c|}{\textbf{Wing Delivery}} & \multicolumn{2}{c|}{\textbf{ADS-B}} & \multicolumn{2}{c|}{\multirow{2}{*}{\textbf{Chat}}} & \multicolumn{2}{c|}{\multirow{2}{*}{\textbf{Other}}}  & \multirow{2}{*}{\textbf{Total}}  \\ 
 & \multicolumn{2}{c|}{} & \multicolumn{2}{c|}{\textbf{Interface}} & \multicolumn{2}{c|}{\textbf{Display}} & \multicolumn{2}{c|}{}& \multicolumn{2}{c|}{} & \\ \hline
\textbf{10:Low} &  434 & 2\%	& 4,529	& 25\%& 7,987	& 44\% & 57	& 0\% & 5,321	& 29\% & 18,328\\ \hline
\textbf{24:Low} & 437	& 3\% & 2,858	& 21\% & 6,166	& 45\% & 9	& 0\% & 4,288	& 31\%& 13,758\\ \hline
\textbf{24:High} & 1,086	& 4\% & 5,741	& 19\% & 16,048	& 54\%& 48	& 0\% & 6,948	& 23\% & 29,871\\ \hline
\textbf{Total} & 1,957	& 3\% & 13,128	& 21\% & 30,201	& 49\% & 114	& 0\% & 16,557	& 27\% & 61,957\\ \hline
\end{tabular}
}
\end{table}

Overall, 72\% of the PICs total fixations were on the four displays for just under 80\% of the fixation duration across the conditions. The remainder of the time, PICs were looking off monitor or on a monitor section not allocated to one of the displays. The fixation counts and durations presented in Tables \ref{Table:FixCountOverall} and \ref{Table:FixDurOverall}, respectively, show that the overall results are similar across the conditions. Recall that 10:Low and 24:Low had three PICs each; thus, the total fixation counts and durations are lower than 24:High that contains results for all PICs. 

\begin{table}[!hbt]
\centering
\caption{The total fixation durations (hh:mm:ss) by trial and area.} 
\label{Table:FixDurOverall}
\begin{tabular}{|c|c|c|c|c|c|c|}
\hline
 \multirow{2}{*}{\textbf{Trial}} & \multirow{2}{*}{\textbf{METAR}} & \textbf{Wing Delivery} & \textbf{ADS-B} & \multirow{2}{*}{\textbf{Chat}} & \multirow{2}{*}{\textbf{Other}} & \multirow{2}{*}{\textbf{Total}} \\ 
 & & \textbf{Interface} & \textbf{Display} & & &\\ \hline
\textbf{10:Low} & 00:04:06	& 00:35:39	& 01:27:00	& 00:00:27	& 00:37:42	& 02:44:54\\ \hline
\textbf{24:Low} &  00:07:18	& 00:32:42	& 01:31:03	& 00:00:09	& 00:34:06	& 02:45:18\\ \hline
\textbf{24:High} & 00:14:12	& 00:57:24	& 03:14:54	& 00:00:36	& 00:58:24	& 05:25:30\\ \hline
\textbf{Total} & 00:25:36	& 02:05:45	& 06:12:57	& 00:01:12	& 02:10:12	& 10:55:42\\\hline
\end{tabular}
\end{table}

Focusing on the four displays, the ADS-B display had 67\% of all fixations for 71\% of the total fixation duration. The Wing delivery interface had 29\% of the fixations for 24\% of the fixation duration, followed by the METAR (fixations: 4\%, fixation duration: 5\%) and the chat (fixations: 0.25\%, fixation duration: 0.23\%). These results were expected given the instructions provided to the PICs. The PICs had no chat responsibilities, other than monitoring the chat. The PICs after the first two PICs during the first week were told the Weather update occurred at 45 minutes $\pm10$ minutes into the trial, which represents the standard operational METAR update period. These overall results align with the experimenter's observations, that the PICs primarily focused on the ADS-B display. Additionally, the PICs indicated that monitoring the ADS-B was critical for deciding whether or not to pause the system. Overall, these results demonstrate that the PICs were engaged in the tasks throughout each trial, and focused their visual attention consistent with successful completion of their job duties. 

\begin{figure}[!htb]
\centering
\begin{subfigure}{0.32\textwidth}
    \centering
    \includegraphics[width=1\linewidth, keepaspectratio]{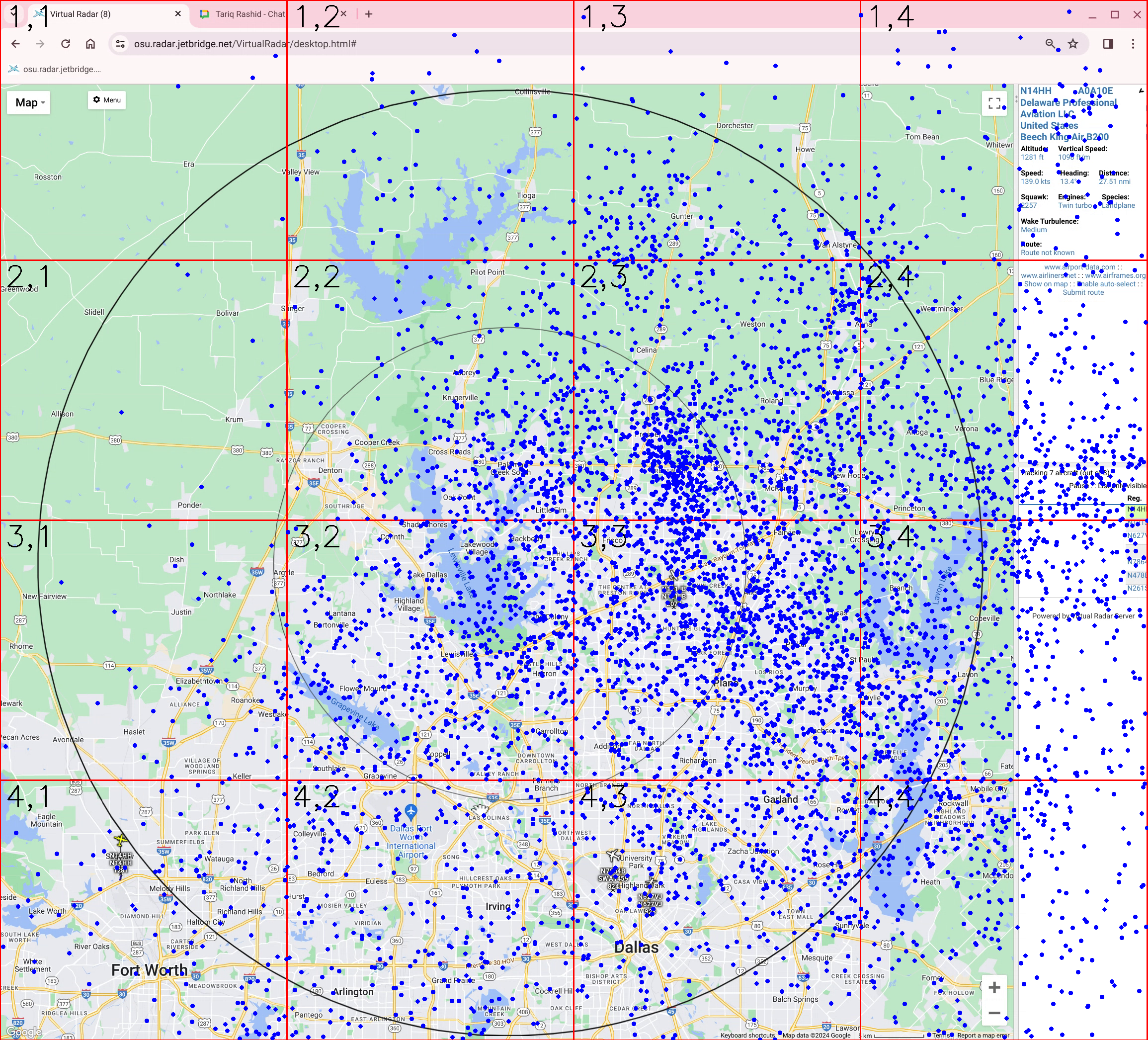}
    \caption{10:Low, three PICs.}
    \label{Fig:FixationsOverall10:Low}
\end{subfigure}
\hfill
\begin{subfigure}{0.32\textwidth}
    \centering  
    \includegraphics[width=1\linewidth, keepaspectratio]{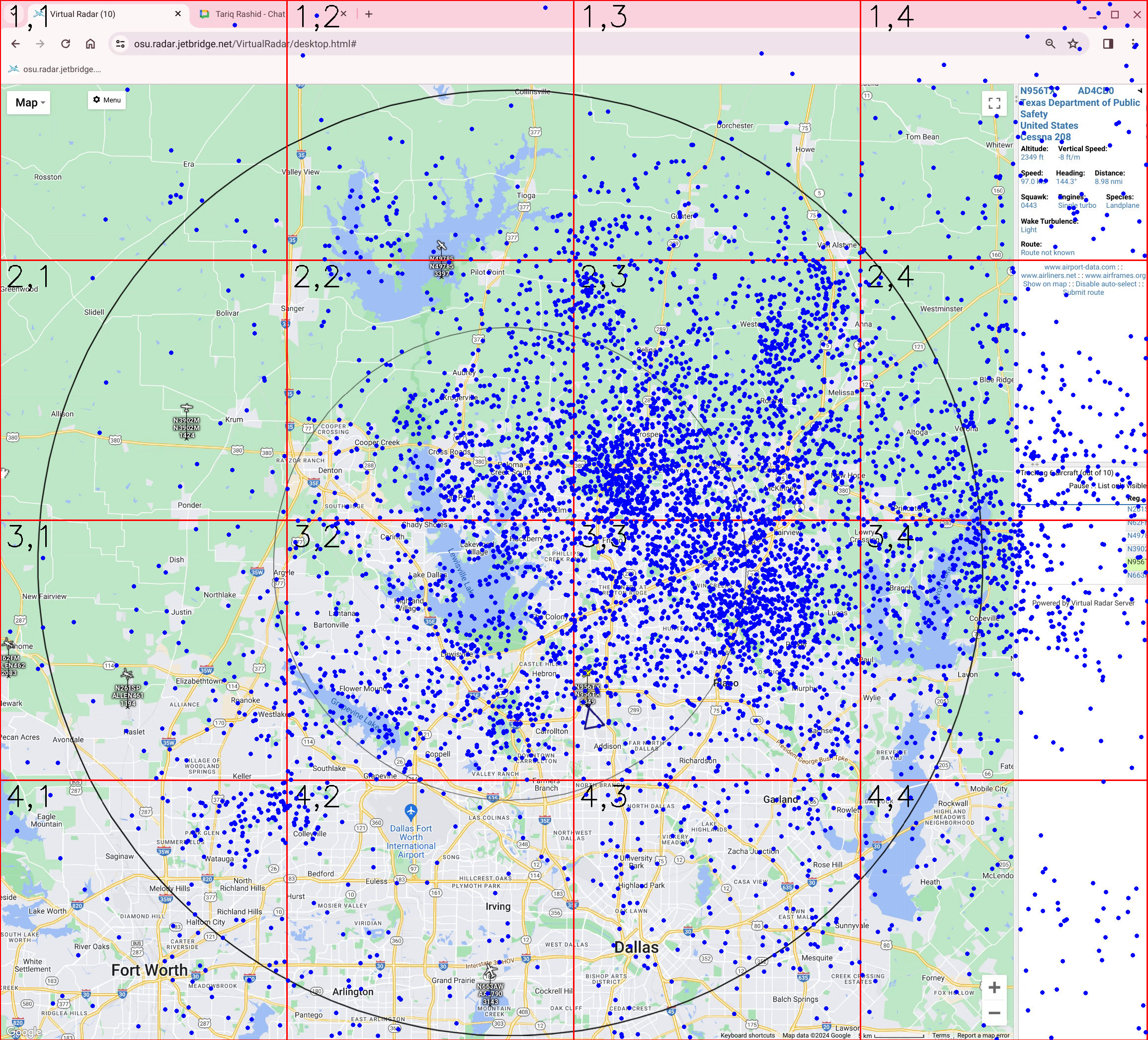}
    \caption{24:Low, three PICs.}
    \label{Fig:FixationsOverall24:Low}
\end{subfigure}
\hfill
\begin{subfigure}{0.32\textwidth}
    \centering  
    \includegraphics[width=1\linewidth, keepaspectratio]{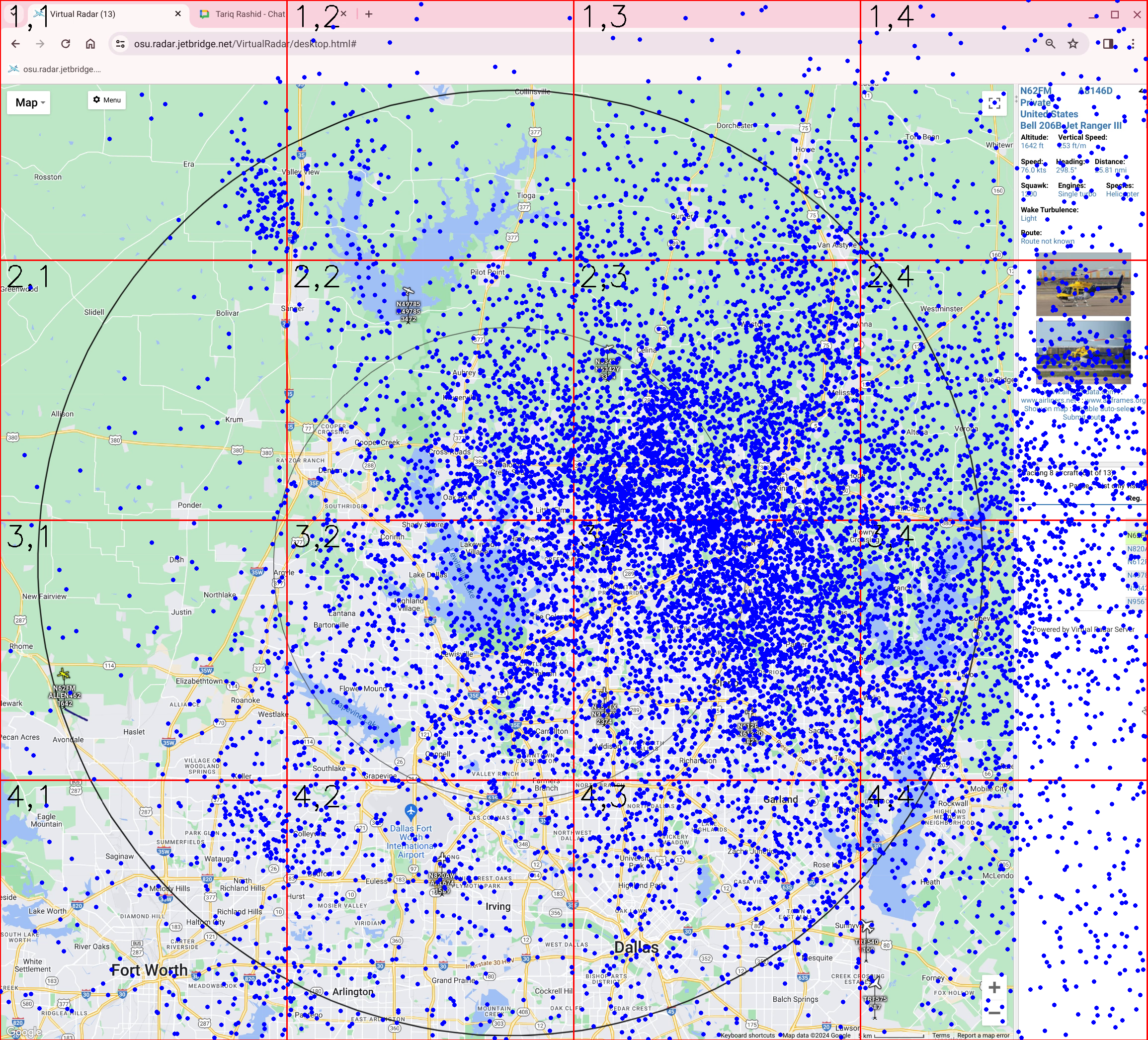}
    \caption{24:High, six PICs.}
    \label{Fig:FixationsOverall24:High}
\end{subfigure}
\caption{The ADS-B display eye tracking fixations mapped to the AOIs by trial. Note that the 24:High plot looks denser because it contains data for six PICs. }
\label{fig:GridFixationsOverall}
\end{figure}

The PICs were allowed to zoom the ADS-B map in or out to their desired viewing level, although the display was originally set identically at the start of each trial. The eye fixations were mapped to the pixels on the ADS-B display. 75\% of all trials were zoomed to roughly the same level, as shown in Figure \ref{fig:GridFixationsOverall}. The ADS-B web-based tool received live crewed aircraft updates, with no log files to provide exact zoom levels for matching fixation pixels to exact map areas. Thus, the fixations were mapped using the underlying map zoomed to the level represented in the majority of the trials. A script was used to leverage the ADS-B tool's API to inject the encounter crewed aircraft for the DAA:Single and DAA:Double tasks. 

The 10:Low trials had the highest number of fixations per PIC, on average 6,109, while the 24:Low trials had the lowest, on average 4,568. The 24:High trials average fixations per PIC was 4,979. While 10:Low had more total fixations than 24:Low, the 10:Low trial fixations are more distributed, as shown in Figures \ref{Fig:FixationsOverall10:Low} and \ref{Fig:FixationsOverall24:Low}, respectively. Note that the 24:High trial plot contains more total fixations given that all six PICs completed the trial. Overall the results demonstrate that the PICs were engaged in completing each trial's tasks. 

\begin{table}[!hbt]
\centering
\caption{The total fixation counts by trial, area, and task. } 
\label{Table:FixCountTasks}
\begin{tabular}{|c|c|c|c|c|c|c|}
\hline
\multirow{2}{*}{\textbf{Trial}} & \multirow{2}{*}{\textbf{METAR}} & \textbf{Wing Delivery} & \textbf{ADS-B} & \multirow{2}{*}{\textbf{Chat}} & \multirow{2}{*}{\textbf{Other}} & \multirow{2}{*}{\textbf{Total}}  \\ 
 & & \textbf{Interface} & \textbf{Display} & & & \\ \hline
 \multicolumn{7}{|c|}{\textbf{Nominal \#1}} \\ \hline
\textbf{10:Low} &  84	& 819	& 1,180	& 6	& 883 & 2,972\\ \hline
\textbf{24:Low} &  42	& 516	& 724	& 2	& 839 & 2,123 \\ \hline
\textbf{24:High} & 207	& 778	& 2,514	& 10	& 976 & 4,485\\ \hline
 \multicolumn{7}{|c|}{\textbf{DAA: Single}} \\ \hline
\textbf{10:Low} & 25	& 910	& 1,475	& 12	& 878 & 3,300\\ \hline
\textbf{24:Low} & 20	& 610	& 1,214 &	1	& 830 & 2,675\\ \hline
\textbf{24:High} & 57	& 1,091	& 2,836	& 8	& 868 & 4,860 \\ \hline
 \multicolumn{7}{|c|}{\textbf{Nominal \#2}} \\ \hline
\textbf{10:Low} &  41	& 817	& 1,205	& 9	& 930 & 3,002\\ \hline
\textbf{24:Low} & 44	& 517	& 1,118	& 2	& 684 & 2,365\\ \hline
\textbf{24:High} & 185	& 838	&2,380	& 8	& 1,160 & 4,571 \\ \hline
 \multicolumn{7}{|c|}{\textbf{DAA: Double}} \\ \hline
\textbf{10:Low} & 27	& 644	& 1,583	& 6	& 703 & 2,963\\ \hline
\textbf{24:Low} &  40	& 414	& 1,160	& 0	& 713 & 2,327\\ \hline
\textbf{24:High} & 86	& 1,352	& 2,976	& 9	& 1,305 & 5,728\\ \hline
 \multicolumn{7}{|c|}{\textbf{Nominal \#3}} \\ \hline
\textbf{10:Low} & 67	& 756	& 1,294	& 17	& 823 & 2,957\\ \hline
\textbf{24:Low} &  46	& 453	& 985	& 1	& 626 & 2,111\\ \hline
\textbf{24:High} & 63	& 751	& 2,637	& 8	& 1,148 & 4,607\\ \hline
 \multicolumn{7}{|c|}{\textbf{Weather}} \\ \hline
\textbf{10:Low} & 190	& 583	& 1,250	& 7	& 1,104 &3,134\\ \hline
\textbf{24:Low} & 245	& 348	& 965	& 3	& 596 & 2,157\\ \hline
\textbf{24:High} & 488	& 931	& 2,705	& 5	& 1,491 & 5,620\\ \hline
\end{tabular}
\end{table}

\begin{table}[!hbt]
\centering
\caption{The total fixation duration (hh:mm:ss) by trial, area, and task. } 
\label{Table:FixDurTasks}
\begin{tabular}{|c|c|c|c|c|c|c|}
\hline
\multirow{2}{*}{\textbf{Trial}} & \multirow{2}{*}{\textbf{METAR}} & \textbf{Wing Delivery} & \textbf{ADS-B} & \multirow{2}{*}{\textbf{Chat}} & \multirow{2}{*}{\textbf{Other}}  & \multirow{2}{*}{\textbf{Total}}  \\ 
 & & \textbf{Interface} & \textbf{Display} & & &\\ \hline
 \multicolumn{7}{|c|}{\textbf{Nominal \#1}} \\ \hline
\textbf{10:Low} & 00:00:06	& 00:05:27	& 00:13:00	& 00:00:04	& 00:06:06	& 00:24:43)\\ \hline
\textbf{24:Low} &  00:01:00	& 00:06:03	& 00:13:45	& 00:00:00	& 00:06:33	& 00:27:21\\ \hline
\textbf{24:High} &00:02:42	& 00:07:00	& 00:29:42	& 00:00:06	& 00:07:36	& 00:47:06\\ \hline
 \multicolumn{7}{|c|}{\textbf{DAA: Single}} \\ \hline
\textbf{10:Low} &  00:00:30	& 00:07:09	& 00:15:03	& 00:00:02	& 00:06:33	& 00:29:17\\ \hline
\textbf{24:Low} &  00:00:30	& 00:05:15	& 00:18:06	& 00:00:03	& 00:06:42	& 00:30:36 \\ \hline
\textbf{24:High} & 00:01:12	& 00:10:54	& 00:33:54	& 00:00:06	& 00:06:36	& 00:52:42 \\ \hline
 \multicolumn{7}{|c|}{\textbf{Nominal \#2}} \\ \hline
\textbf{10:Low} &  00:00:09	& 00:06:51	& 00:13:30	& 00:00:03	& 00:06:45	& 00:27:18\\ \hline
\textbf{24:Low} & 00:01:00	& 00:06:00	& 00:15:18	& 00:00:03	& 00:05:33	& 00:27:54\\ \hline
\textbf{24:High} & 00:02:06	& 00:10:42	& 00:29:24	& 00:00:06	& 00:10:06	& 00:52:24 \\ \hline
 \multicolumn{7}{|c|}{\textbf{DAA: Double}} \\ \hline
\textbf{10:Low} & 00:00:57	& 00:05:39	& 00:16:48	& 00:00:03	& 00:04:48	& 00:28:15\\ \hline
\textbf{24:Low} & 00:00:54	& 00:05:15	& 00:15:33	& 00:00:00	& 00:06:21	& 00:28:03)\\ \hline
\textbf{24:High} &00:01:18	& 00:12:06	& 00:35:54	& 00:00:06	& 00:12:18	& 01:01:42\\ \hline
 \multicolumn{7}{|c|}{\textbf{Nominal \#3}} \\ \hline
\textbf{10:Low} & 00:00:30	& 00:06:21	& 00:14:36	& 00:00:09	& 00:05:24	& 00:27:00 \\ \hline
\textbf{24:Low} &  00:01:03	& 00:05:57	& 00:14:21	& 00:00:00	& 00:04:54	& 00:26:15 \\ \hline
\textbf{24:High} &00:01:12	& 00:07:48	& 00:32:48	& 00:00:06	& 00:08:18	& 00:50:12\\ \hline
 \multicolumn{7}{|c|}{\textbf{Weather}} \\ \hline
\textbf{10:Low} &  00:01:54	& 00:04:12	& 00:14:03	& 00:00:06	& 00:08:06	& 00:28:21 \\ \hline
\textbf{24:Low} & 00:02:51	& 00:04:12	& 00:14:00	& 00:00:03	& 00:04:03	& 00:25:09 \\ \hline
\textbf{24:High} & 00:05:42	& 00:08:54	& 00:33:12	& 00:00:06	& 00:13:30	& 1:01:24) \\ \hline
\end{tabular}
\end{table}

Overall, 69\% of fixations on the ADS-B display focused on AOIs 2,3 - 2,4 and 3,3 - 3,4 for 76\% of the total fixation duration, see Figure \ref{fig:GridCells} for AOI layout. AOIs 2,3 and 3,3 had the highest fixation counts and durations. The ADS-B display's left side (i.e.,\ AOIs 1,1 - 4,1) received the lowest focus of attention across all trials and tasks, $< 2\%$ of total fixations for $\leq 1\%$ of the duration (i.e.,\ 3 minutes), with the top row (i.e.,\ AOIs 1,1 - 1,4) being the second lowest, with $\leq 10\%$ for 6\% of the total fixation duration. The 10:Low trial fixations were slightly more dispersed across the AOIs, but the fixation durations did not differ from the other trials. Overall, the X:Low trials had slightly more fixations and for slightly longer durations in AOIs 2,2 and 3,2 as compared to the 24:High trial. 

The X:Low Nominal tasks had 15,530 total fixations for a total duration of 02:40:31, exceeding the 24:High Nominal tasks' fixations (i.e., 13,663) and duration (i.e.,\ 02:29:42), details in Tables \ref{Table:FixCountTasks} and \ref{Table:FixDurTasks}. 
24:Low had 194 more total fixations for a 00:02:29 longer duration. Thus, only very small differences in these overall metrics existed across the Nominal tasks within any particular trial. 

The Nominal task's fixation distribution across the ADS-B AOIs were similar to those presented in Figure \ref{fig:GridFixationsOverall}. The highest ADS-B Nominal task fixations occurred in AOIs 2,3 - 2,4 and 3,3 - 3,4 with the highest durations irrespective of trial. Across the Nominal tasks and trials, AOI 2,3 had the highest number of fixations for the longest duration. Specific to the Nominal \#1 task and the X:Low trials, the number of fixations and the durations were more evenly distributed across the 2,3 - 2,4 and 3,3 - 3,4 AOIs, whereas the X:Low Nominal \#2 and \#3 trials and all 24:High Nominal trials had a higher focus on AOIs 2,3 and 3,3. 

\subsubsection{Number of Nests}

Across the X:Low trials, the number of nests did not induce higher fixation levels. The 24:Low trials had 25\% fewer total fixations with a slightly longer 0.24\% total fixation duration. The 24:Low PICs focused more on the ADS-B and Weather displays than the 10:Low PICs. The increased Weather display focus is likely due to session briefings emphasizing the METAR updates. The 24:Low total fixation counts fell the most (-37\%) for the Wing interface, while the ADS-B total fixations fell by 23\%. This difference may be due to the spatial distribution of the 24:Low trial's nests; however, it must be noted that 24:Low's fewer fixations are less dispersed on the ADS-B display than the 10:Low trial's fixations, as shown in Figures \ref{Fig:FixationsOverall10:Low} and \ref{Fig:FixationsOverall24:Low}. The ADS-B AOI analysis found 74\% of the 24:Low fixations were in AOIs 2,3 - 2,4 and 3,3 - 3,4 for 79\% of the trial's total fixation duration, as compared to 58\% of the 10:Low trial's fixations in those same areas for 69\% of the total fixation duration. 

The Nominal tasks resulted in a very small increase in the total fixation duration during the 24:Low trials as compared to 10:Low, with fewer overall fixations in the 24:Low trials. These small differences are within the margin of error and demonstrate no effect on viewing patterns as a result of increasing the number of nests.

\subsubsection{Number of Active UAS}
Increasing the number of active UAS had no positive impact on either the total number of fixations or fixation durations. Overall the combined X:Low trials had more total fixations (52\%) with a 50.36\% higher total fixation duration as compared to the 24:High trials. As a reminder, each X:Low trial in Tables \ref{Table:FixCountOverall} and \ref{Table:FixDurOverall} reports results for three PICs and 24:High reports results for six PICs. 

The Nominal tasks during X:Low trials had substantially higher total fixations and fixation duration relative to the 24:High trials, as shown in Tables \ref{Table:FixCountTasks} and \ref{Table:FixDurTasks}, respectively. Increasing the number of active UAS did not increase these metrics during the Nominal tasks. 

The PICs focused their attention differently with an increased number of active UAS. The X:Low trials had 7\% more total fixations that were not on the four interface displays for a 4\% longer duration. During the X:Low trials the Wing interface had 4\% more total fixations for a 3\% longer duration. The 24:High fixations shifted primarily to the ADS-B and the METAR displays. The 24:High trials focus on the METAR display, in terms of both total fixations and fixation duration, increased by 1\%, which is likely due to learning effects. The ADS-B display focus increased 10\% in total fixations with a 6\% longer duration during the 24:High trials. 
During the 24:High trials, PICs specifically focused more on the four primary AOIs (i.e., 2,3 - 2,4 and 3,3 - 3,4), with a higher fixation count percentage (X:Low 65\%, 24:High 73\%) and a longer duration (X:Low 74\%, 24:High 79\%) than the X:Low trials. 

\subsubsection{Unexpected Events}

Each DAA:X crewed aircraft encounter appeared on the ADS-B display for at least six minutes. These encounters were intended to require the PICs' attention, but not result in pausing operations. Further, it was anticipated that fixations will increase in the AOIs in which the crewed aircraft encounters occurred, since the PICs were trained to track these vehicles. This tracking is necessary to determine if a crewed vehicle encounter met the criteria for pausing operations. 
The Weather updated at approximately minute 52, or 2 minutes into the Weather task. The adverse weather applied to the entire operational area. The updated weather conditions were expected to cause the PIC to pause operations for the ``remainder of the day'', which in this case was the remainder of the trial. The Weather task update was expected to increase the fixation counts and duration on the METAR display. 

\begin{figure}[!htb]
\centering
\begin{subfigure}{0.4\textwidth}
    \centering
    \includegraphics[width=1\linewidth, keepaspectratio]{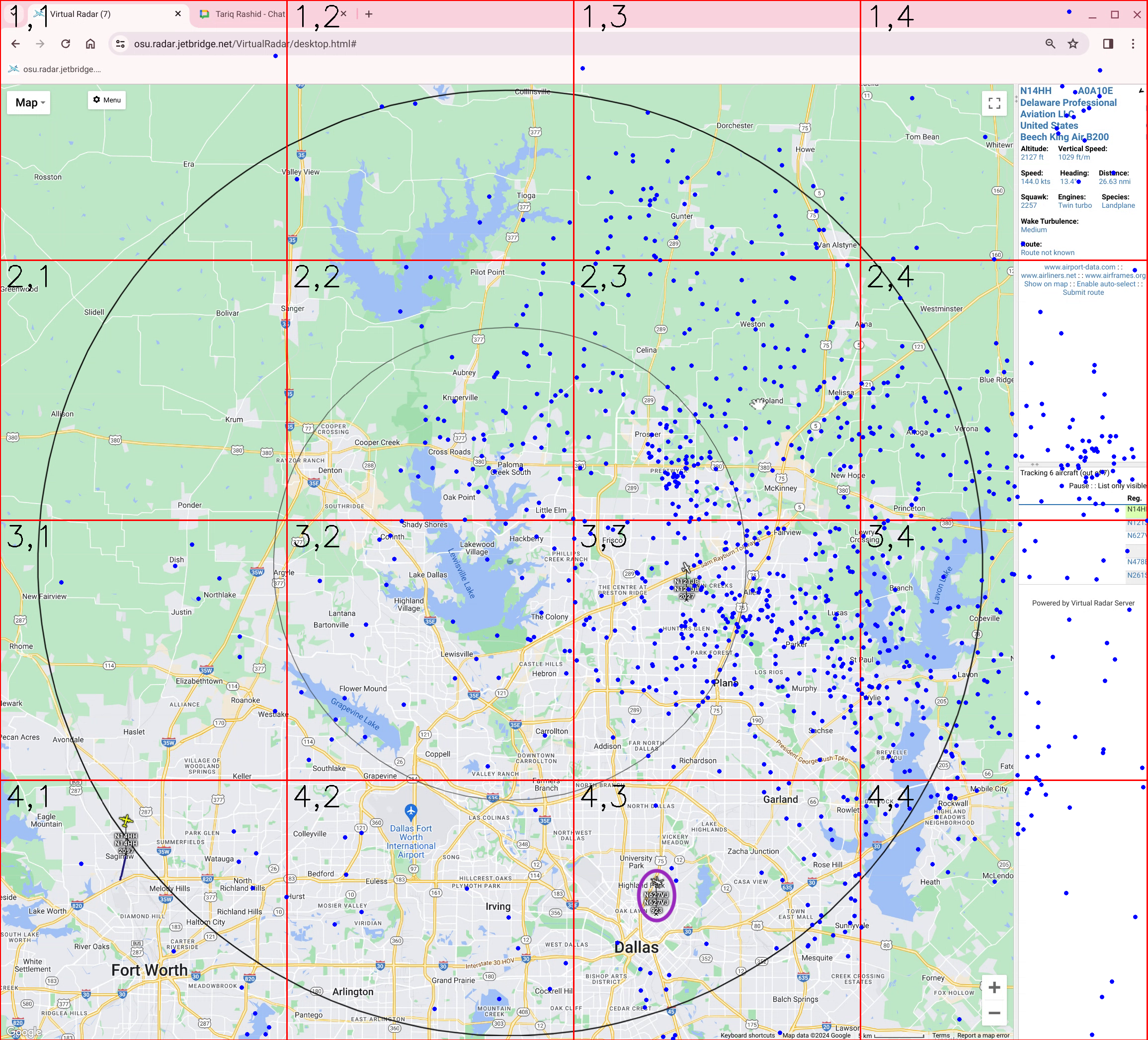}
    \caption{DAA:Single 10:Low trials.}
    \label{Fig:Fix10:LowDAA:Single}
\end{subfigure}
\begin{subfigure}{0.4\textwidth}
    \centering
    \includegraphics[width=1\linewidth, keepaspectratio]{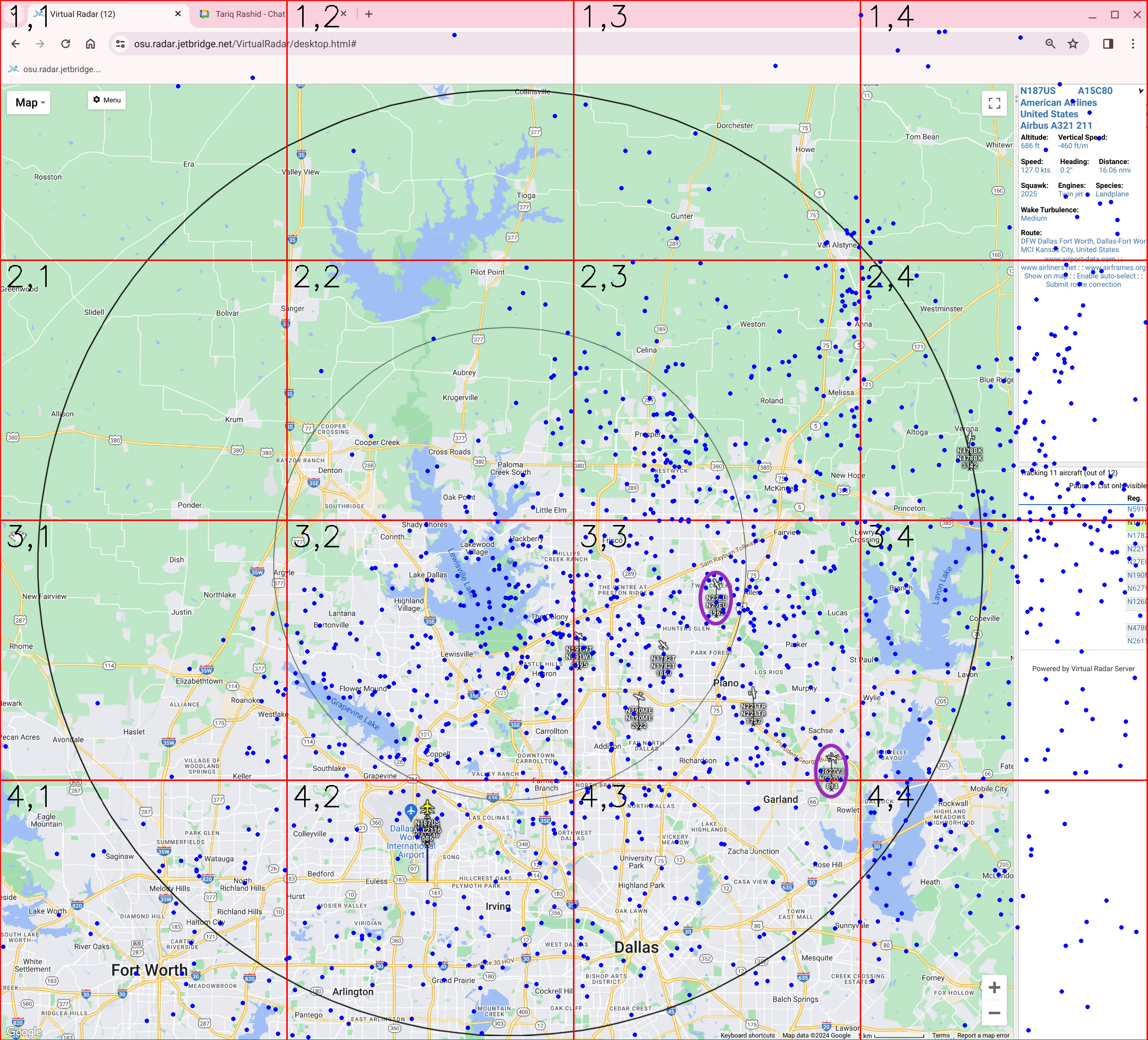}
    \caption{DAA:Double 10:Low trials.}
    \label{Fig:Fix10:LowDAA:Double}
\end{subfigure}
\hfill
\begin{subfigure}{0.4\textwidth}
    \centering  
    \includegraphics[width=1\linewidth, keepaspectratio]{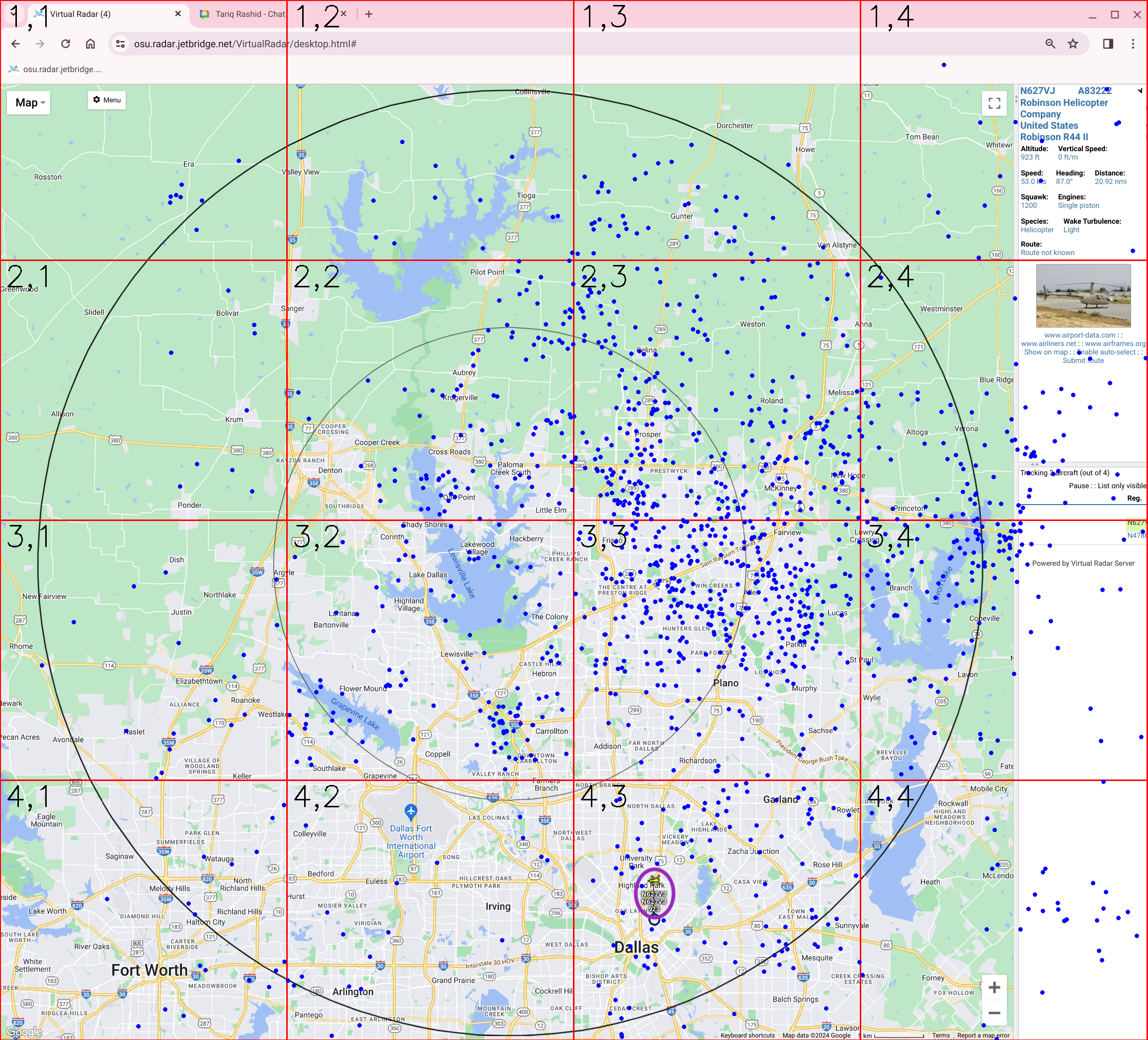}
    \caption{DAA:Single 24:Low trials.}
    \label{Fig:Fix24:LowDAA:Single}
\end{subfigure}
\begin{subfigure}{0.4\textwidth}
    \centering  
    \includegraphics[width=1\linewidth, keepaspectratio]{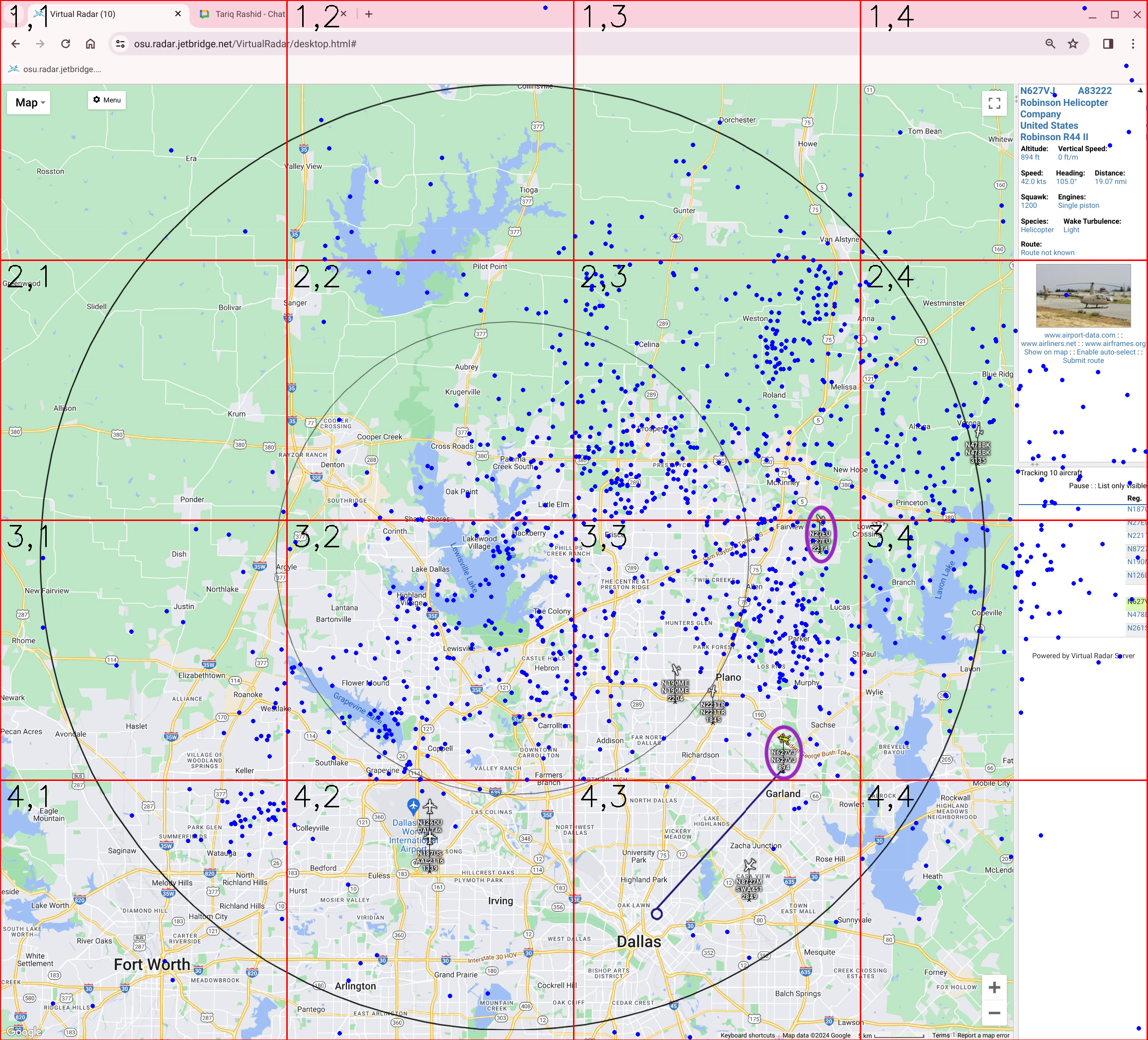}
    \caption{DAA:Double 24:Low trials.}
    \label{Fig:Fix24:LowDAA:Double}
\end{subfigure}
\hfill
\begin{subfigure}{0.4\textwidth}
    \centering  
    \includegraphics[width=1\linewidth, keepaspectratio]{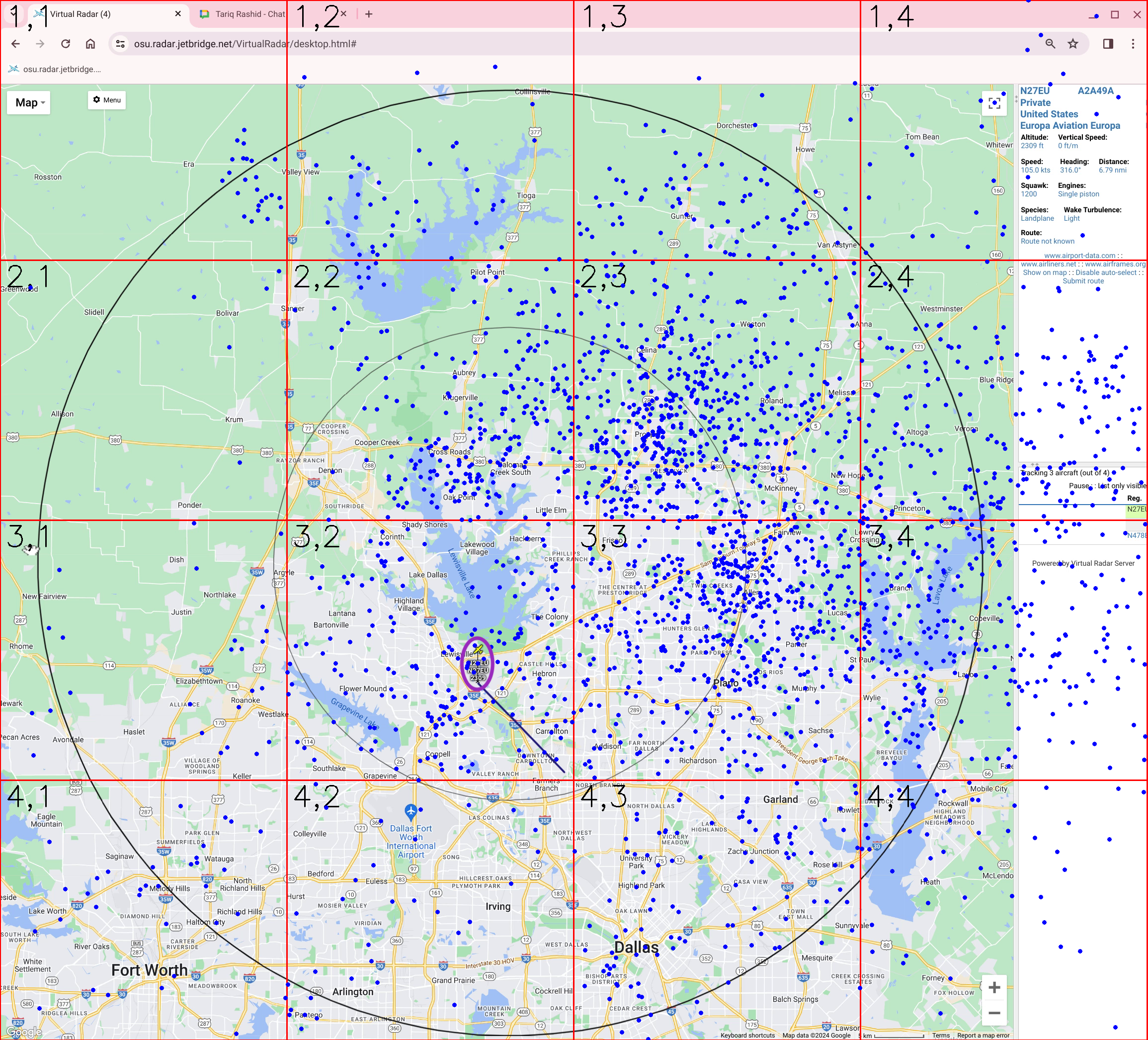}
    \caption{DAA:Single 24:High trials.}
    \label{Fig:Fix24:HighDAA:Single}
\end{subfigure}
\begin{subfigure}{0.4\textwidth}
    \centering  
    \includegraphics[width=1\linewidth, keepaspectratio]{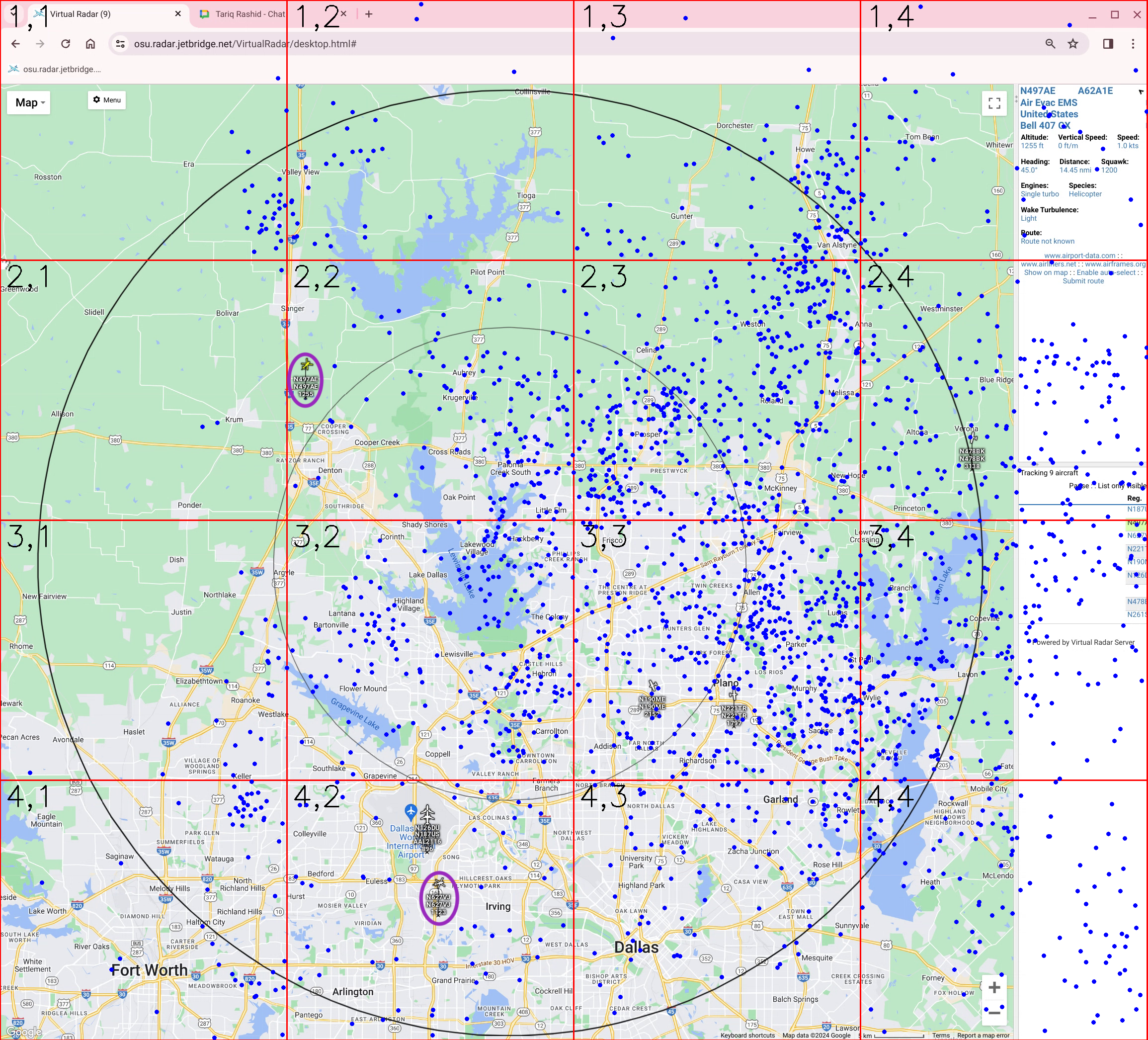}
    \caption{DAA:Double 24:High trials.}
    \label{Fig:Fix24:HighDAA:Double}
\end{subfigure}
\caption{The DAA:X tasks fixations mapped to the AOIs by trial. The purple ovals indicate the locations of the crewed encounter aircraft.}
\label{fig:GridDAATrials}
\end{figure}

\paragraph{Detect And Avoid: Single Crewed Aircraft}
The DAA:Single tasks for the X:Low trials shared the same crewed encounter aircraft, while the 24:High trial had a different crewed encounter aircraft, as shown in Figure \ref{fig:GridDAATrials}. The figures have a purple highlight around the encounter aircraft, and the underlying maps were captured 15 seconds after the encounter crewed aircraft appeared. Variance in the exact start time of the DAA script (see Table \ref{Table:DAATiming}) existed. During all task trials, the encounter crewed aircraft generally remained within the identified AOI. 

The X:Low DAA:Single tasks had the highest total fixations (5,975) for a total 00:59:53 duration, which were higher than the 24:High results (fixations:\ 4,860, duration:\ 00:52:42). 
The 24:High trial's increase in active UAS did not increase the total fixation count or duration, which were ranked third highest across all tasks within the trial. 

The ADS-B remained the most dominant display, followed by the Wing user interface. The X:Low ADS-B fixations were relatively the same across the trials with an 8\% longer duration during the 24:Low trials. The total number of ADS-B display fixations during the 24:High trials increased by 13\%, which had the longest duration (00:33:54) across the DAA:Single task trials. The total X:Low ADS-B fixation duration increased by 24\% during the DAA:Single task as compared to the Nominal \#1 task. While the total 24:High DAA:Single task fixation duration was lower than the X:Low trials, it was 9\% higher than the 24:High Nominal \#1 task's duration. The X:Low DAA:Single encounters occurred, as shown in Figures \ref{Fig:Fix10:LowDAA:Single} and \ref{Fig:Fix24:LowDAA:Single}, in AOI 4,3, with the encounter crewed aircraft roughly remaining in that AOI. There were no increases in the number of fixations on that specific AOI, but there was a 28\% fixation duration increase for that AOI during the 24:Low DAA:Single task as compared to the 10:Low trial. Interestingly, the X:Low DAA:Single task's number of fixations in AOIs 3,3 and 4,3 more than doubled (i.e., a 102\% increase) with a 91\% increase in total duration as compared to the Nominal \#1 task. A smaller, but substantial increase in total fixations (35\%) with a 4\% decrease in duration existed for the adjacent 3,4 and 4,4 AOIs. 
The 24:High crewed aircraft encounter occurred in AOI 3,2, an area that nominally was not a focus of attention. This encounter resulted in a 200\% increase in fixations on that AOI as compared to the Nominal \#1 task, with a 176\% increase in fixations on AOI 2,2 and a 93\% increase for the adjacent 2,3 and 3,3 AOIs. A 322\% increase in fixation durations on AOI 2,2 existed as compared to the Nominal \#1 task, with a 202\% increase for the adjacent 2,2 AOI and 105\% increase for the adjacent 2,3 and 3,3 AOIs. 

\paragraph{Detect And Avoid: Double Crewed Aircraft}
The DAA:Double task's first crewed vehicle encounter timing matched the DAA:Single task, and was followed by the second encounter. The locations of both encounters are highlighted in purple for each condition in Figure \ref{fig:GridDAATrials}, where the underlying maps were captured 15 seconds after the second encounter occurred. Variance in the exact DAA script start times existed. The encounter crewed aircraft generally remained within the indicated AOI. 

There were 5,290 total fixations during the X:Low DAA:Double task for a 00:56:18 total fixation duration (see Tables \ref{Table:FixCountTasks} and \ref{Table:FixDurTasks}, respectively); a decrease from the X:Low DAA:Single task. The 24:High DAA:Double task had more fixations (5,728) for a longer total duration (01:01:42), which is a substantial increase (fixations: 18\%, duration: 17\%) over the same condition's DAA:Single task. This increase is not directly due to the increase in number of active UAS, but more likely a result of the spatial distances between the two crewed encounter aircraft during the 24:High trial. The encounter crewed aircraft during the 24:High trial were separated by a substantial spatial distance as compared to the X:Low DAA:Double trials, as shown in Figures \ref{Fig:Fix10:LowDAA:Double} and \ref{Fig:Fix24:LowDAA:Double} for the X:Low trials and Figure \ref{Fig:Fix24:HighDAA:Double}. The X:Low trials reduced spatial separation between the encounter aircraft permitted gathering information about both aircraft simultaneously, which was not feasible for the 24:High trial.

During the DAA:Double X:Low trials, there were fewer fixations with slightly lower total duration on the Wing user interface display as compared to the DAA:Single task; however, the same comparison for the 24:High trials found a 24\% increase in fixation count and 11\% increase in total duration. The ADS-B display difference during the X:Low trials was a 2\% increase in fixation counts with a corresponding 6\% duration decrease. The 24:High trial's ADS-B display total fixations increased by 49\% with a 6\% increased duration. 

Prior to discussing the DAA:Double results by AOI, note that during the X:Low Nominal \#2 tasks, the fixation counts in AOIs 2,3 and 3,3 were as high or higher than the same AOIs during the DAA:Single task. All surrounding AOIs had lower fixation counts during the Nominal \#2 task as compared to the DAA:Single task. 
The fixation durations had the same pattern by AOI. This result may have occurred because PICs did not know that each task was 10 minutes in length, and may have anticipated similar crewed aircraft encounters in those AOIs during the Nominal \#2 task. 

The X:Low DAA:Double encounters occurred in AOI 3,3 in close proximity to AOIs 2,3 and 4,3. The 10:Low trial's DAA:Double AOI 3,3 had the highest fixation count, which was 49\% higher than the same AOI during the Nominal \#2 task. There were also substantial fixation increases in the adjacent 3,2 (181\%), 4,2 (186\%) and 4,3 (144\%) AOIs. The 24:Low trials had a decrease in fixations for AOIs 2,3 (-20\%) and 3,3 (-7\%), but a substantial increase for AOIs 3,2 (332\%) and 2,4 (83\%), with small increases for the other adjacent AOIs as compared to the Nominal \#2 task. The fixation duration results had similar increases and decreases by AOI. Given the proximity of the crewed encounter aircraft to the AOI boundaries, these differences between the X:Low trials were expected. Overall, the PICs were focused on the AOI in which the encounters occurred. 

The 24:High fixations in AOIs 2,2, 2,3 and 3,2 fell during Nominal \#2 task and increased for AOIs 3,3, 4,2, 4,3 and 4,4, while the remaining adjacent AOIs were relatively unchanged. The Nominal \#2 task durations followed a similar pattern. 
As compared to the Nominal \#2 task, a 144\% increase in fixations and an 85\% increase in fixation duration in AOI 3,2 existed, with slight decreases in fixation count for AOIs 2,2 and 4,2 during the 24:High DAA:Double task. There was a 4\% increase in fixation duration for AOI 2,2 and a 73\% increase in AOI 4,2. An important difference during this task, as compared to all other tasks across trials is a 215\% increase in fixation counts and a 532\% increase in duration for AOI 1,4 where the aircraft information in the display's upper right corner resides. This increase is likely due to information gathering regarding the crewed aircraft, as PICs were attempting to further deduce the nature of the encounter crewed aircraft to help guide subsequent actions.  This is supported by the high amount of ADS-B interactions in the 24:High trial.  

\paragraph{Weather}
The X:Low Weather tasks resulted in a fixation count of 5,291 for a total duration of 00:53:30, which was lower than the 24:High trial's 5,620 total fixations with a 01:01:24 total duration. Given the larger number of active UAS during the 24:High trial, this result is not surprising, as it takes longer for the active UAS to land upon mission completion. 

The total Weather display fixations and associated durations was very low compared to the ADS-B and Wing interface displays. Generally, during the Weather task, the PICs substantially increased their fixations on the Weather display as compared to the same display during other tasks. 
The X:Low trials resulted in fewer fixations (10:Low 2\%, 24:Low 3\%) of the respective trial's total fixations, while the 24:High trial fixations were 4\% of the trial's total (see Table \ref{Table:FixCountTasks}). The durations (see Table \ref{Table:FixDurTasks}) were similarly small across the trial. During this task, the number of ADS-B display fixations and the total duration remained relatively the same as compared to the other tasks, irrespective of trial. The distribution of the ADS-B fixations across the AOIs were representative of the overall results provided in Figure \ref{fig:GridFixationsOverall}. Overall, the fixation AOIs and durations on each AOI were representative of the overall AOI results. The increased focus on the Weather display generally resulted in less focus on the Wing user interface across all trials. 

\subsection{Workload}

\begin{figure}[htb]
\centering
\begin{subfigure}{0.99\textwidth}
    \centering
    \includegraphics[width=1\linewidth, keepaspectratio]{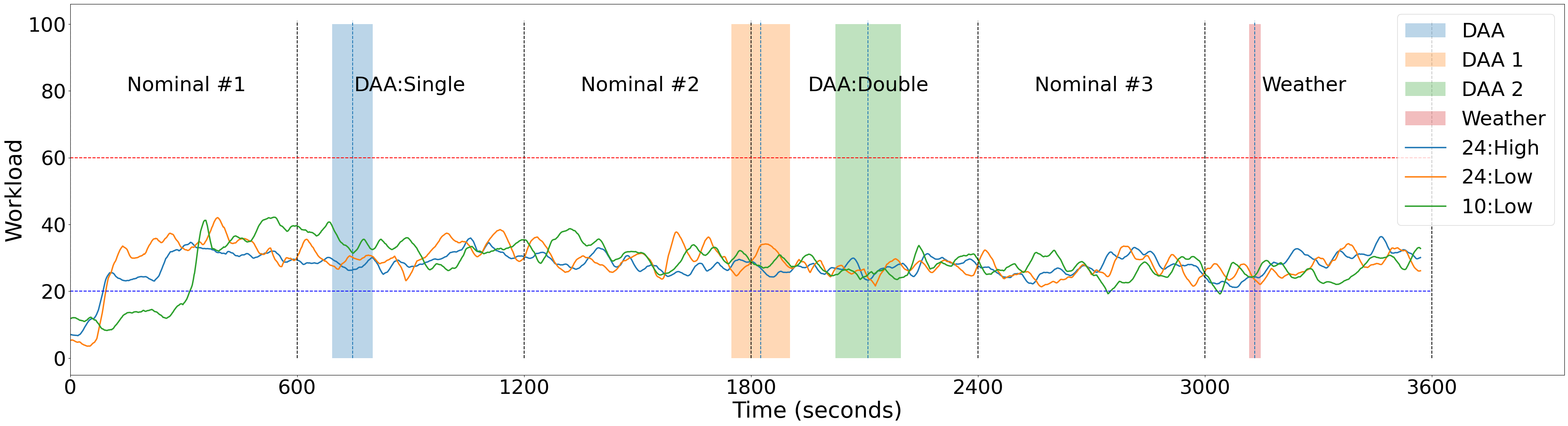}
    \caption{All trials.}
    \label{Fig:OverallWorkload-Trial}
\end{subfigure}
\hfill
\begin{subfigure}{0.99\textwidth}
    \centering  
    \includegraphics[width=1\linewidth, keepaspectratio]{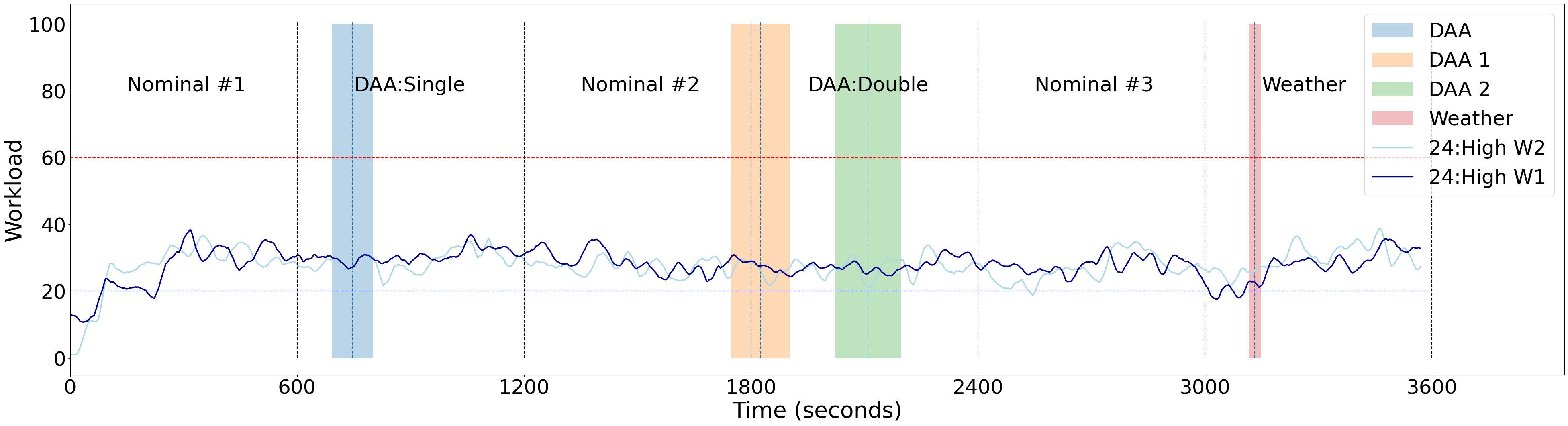}
    \caption{The 24:High trials by week.}
    \label{Fig:OverallWorkload-Trial2Week}
\end{subfigure}
\caption{The mean overall workload estimates by tasks for (a) all trials, and (b) by week (W1 or W2) for the 24:High trials. The gray bars represent transitions between tasks, and colored bars represent the mean and std timing of unexpected events. }
\label{fig:figures}
\end{figure}

The workload results include the physiological estimates of overall workload and each workload component (i.e.,\ cognitive, visual, speech, auditory, gross motor, fine motor and tactile) as well as the subjective workload responses collected at minutes three and six during each 10m task. Across all three trials, there was virtually no difference in the mean overall workload estimates, as shown in Table \ref{Table:OverallWorkload}. The individual workload estimates were generally in the normal workload range. The estimates were in the underload range (i.e.,\ $< 20$) during the first Nominal task's ramp up period, shown in Figure \ref{Fig:OverallWorkload-Trial}, and briefly during the third Nominal and Weather conditions for the 24:High condition. The overall workload estimates never approached the overload threshold (i.e.,\ $>60$) for any condition. 

\begin{table}[hbt]
\centering
\caption{The overall workload estimate descriptive statistics by trial. 
} 
\label{Table:OverallWorkload}
\begin{tabular}{|c|c|c|}
\hline
\textbf{10:Low} & \textbf{24:Low} & \textbf{24:High} \\ \hline
 28.51 (7.64) & 29.39 (7.11) & 29.21 (6.83) \\ \hline
\end{tabular}
\end{table}

The number of active UAS during the 24:High trial varied by week, with week two having substantially fewer active UAS. The 24:High trial overall workload estimates by week were examined to ensure that the discrepancy in active UAS did not create bias in those estimates. Despite the difference in the number of active UAS by week for this condition, as shown in Figure \ref{Fig:OverallWorkload-Trial2Week}, there was not a substantial difference in the estimated workload. Thus, the remainder of the 24:High results are collapsed across data collection weeks. 

\begin{table}[hbt]
\caption{The overall workload estimates' descriptive statistics by trial and task. } 
\label{Table:WorkloadTasks}
\resizebox{\columnwidth}{!}{
\begin{tabular}{|l|c|c|c|c|c|c|}
\hline
\multirow{2}{*}{\textbf{Trial}} & \textbf{Nominal} & \textbf{DAA:} & \textbf{Nominal} & \textbf{DAA:}  & \textbf{Nominal} & \multirow{2}{*}{\textbf{Weather}} \\ 
 & \textbf{\#1} & \textbf{Single} & \textbf{\#2} & \textbf{Double}  & \textbf{\#3} & \\ \hline
10:Low  &  31.65 (6.85) & 30.40 (6.35) & 26.67 (7.13) & 26.18 (6.21) &  25.25 (7.00) & 28.70 (7.86)   \\ \hline
24:Low & 34.22 (5.55) & 32.73 (6.58) & 27.63 (4.92) & 28.34 (5.53) & 24.53 (7.75) & 28.20 (6.25) \\ \hline
24:High & 30.65 (6.66) & 30.41 (6.18) & 28.60 (6.20) & 27.21 (6.08) & 27.69 (6.25) & 30.46 (7.24)  \\ \hline
\end{tabular}
}
\end{table}

The mean overall workload estimates tended to be highest during the first Nominal task (Nominal \#1), with mostly slight decreases until the Weather task across all trials, as shown in Table \ref{Table:WorkloadTasks}. Many of the PICs explicitly commented that they felt their workload was higher during the Nominal \#1 task, because they were unsure what to expect from the trials, especially during their very first trial (i.e.,\ either 10:Low or 24:Low). While the overall Nominal workload results for the X:Low trials are higher than the 24:High trial, they are all within one standard deviation and are not generally different. 

The Nominal tasks all had generally the same overall workload estimate means. A clear trend was that the overall workload decreased with each successive Nominal task (\#1 - \#2 - \#3). The 24:High trial had the lowest Nominal \#1 task mean overall workload estimate, with the smallest decline, 2.96, across the Nominal tasks. The 24:Low trial's Nominal task's mean overall workload estimate dropped the most from the first to the third, 9.69. 

The highest mean overall workload estimates during the unexpected events occurred primarily for the DAA:Single task across all three trials, and for the 24:High trial Weather event. The DAA:Double task's overall mean workload did not substantially differ from the Nominal tasks, or the Weather event. 

\begin{table}[hbt]
\caption{Each workload components' workload estimates' descriptive statistics by trial. The Gross Motor workload values were zero for all trials.} 
\label{Table:WorkloadComponents}
\begin{tabular}{|l|c|c|c|c|c|c|}
\hline
\multirow{2}{*}{\textbf{Trial}} & \multirow{2}{*}{\textbf{Cognitive}} & \multirow{2}{*}{\textbf{Visual}} & \multirow{2}{*}{\textbf{Speech}} & \multirow{2}{*}{\textbf{Auditory}}  & \textbf{Fine} & \multirow{2}{*}{\textbf{Tactile}} \\ 
 & & & & & \textbf{Motor} &\\ \hline
10:Low  &  9.47 (3.96) & 10.91 (2.71)  & 0.78 (1.35)  & 1.11 (1.31) &   3.02 (1.19) &3.22 (1.28)    \\ \hline
24:Low  & 10.01 (4.02) & 11.14 (2.62) & 0.63 (1.24)  & 1.15 (1.29) & 3.13 (1.13) & 3.32 (1.23)  \\ \hline
24:High & 9.99 (3.60) & 11.17 (2.53) & 0.58 (1.20) & 1.14 (1.30)   & 3.13 (1.04) & 3.19 (1.16)   \\ \hline
\end{tabular}
\end{table}

The expectation was that the PIC's task is primarily a visual task with a heavy cognitive workload element. Both fine motor and tactile workload were also expected to have a reasonable level of workload based on moving the mouse and pressing mouse buttons, respectively. The speech and auditory workload were related to responding to the in situ subjective workload and SA probes. The workload component estimates matched the expectations, as shown in Table \ref{Table:WorkloadComponents}. The table shows that within a particular component there was very little variability, with all such variances being within one standard deviation. Given these results, the individual workload component estimates are not discussed further.  

\subsubsection{Number of Nests}

The impact of the number of nests on the PIC's workload performance was investigated. Trials 10:Low and 24:Low increased the number of nests from ten to twenty-four, respectively. The large number of nests resulted in an exceptionally small increase in the mean estimated overall workload, as shown in Table \ref{Table:OverallWorkload} and Figure \ref{Fig:OverallWorkload-Trial}.  This change in workload for a larger number of nests was not statistically reliable. 
When the number of nests were compared across the Nominal tasks for both 10:Low and 24:Low trials, the results indicated that the number of nests did not produce a reliable change in overall workload over the minutes or seconds of these Nominal trials.
These results confirmed that under Nominal conditions, the number of nests did not produce any noticeable impact on overall workload.

\subsubsection{Number of Active UAS}

A commonly accepted hypothesis is that the number of active UAS impacts PICs' workload. While the mean number of UAS more than doubled from the X:Low trials, the 24:High mean overall workload did not change consistent with this increase.  Statistical examination of overall workload for the Nominal tasks in the X:Low and 24:High trials confirmed that increasing the number of UAS did not change workload across the minutes or seconds of these tasks (shown in Figure \ref{Fig:OverallWorkload-Trial}).
These results suggest that increasing UAS under Nominal conditions did not impact the amount of overall workload experienced by the PICs.

\subsubsection{Unexpected Events}

It was anticipated that PIC workload increases during the unexpected events tasks. Specifically, the DAA events were designed with the intention of raising workload, but also to mitigate the likelihood that a PIC will pause operations. When operations are paused, no new UAS become active, and as the UAS complete their deliveries, they return to a nest and land. Thus, an extended pause can substantially reduce the number of active UAS, although deliveries can queue to become active once operations are resumed. 
The adverse Weather update was intended to result in the PIC pausing operations through the trial completion. 

\paragraph{Detect And Avoid: Single Crewed Aircraft}

The DAA:Single task generally had higher overall estimated mean workload than any other 10m task period (aside from Nominal \#1), as shown in Table \ref{Table:WorkloadTasks} and Figure \ref{Fig:OverallWorkload-Trial}.  While overall workload estimates were higher during these DAA tasks than in the Nominal control tasks when comparing the 10:Low to 24:Low trials ($F(1,4) = 22.25, p = .01, \eta^2 = .26$), the DAA:Single tasks did not interact with the number of nests to change workload across minutes or seconds of the task. 
When considering the number of active UAS, the presence of the DAA:Single encounter produced an overall slight elevation of workload relative to the Nominal tasks ($F(1,5) = 33.35, p < .01, \eta^2 = .13$); however, this change was consistent in magnitude across the X:Low and 24:High conditions. Further, this change in the number of active UAS did not produce a reliable change in the estimated overall workload across minutes and seconds of the task. 
These results demonstrate that while the DAA:Single task did modestly increase workload, this increase in workload did not at all change with more nests or more active UAS, and was stable across the manipulated conditions.

\paragraph{Detect And Avoid: Double Crewed Aircraft}
Unexpectedly, the two crewed aircraft encounter (i.e., DAA:Double) task resulted in the \textit{lowest} overall estimated workload means; however, this decrease was not reliably different from the estimated overall workload for the Nominal tasks. 
When considering the increase in the number of nests, there did appear to be a modest increase in the estimated overall workload $(F(531,2124) = 1.63, p < .001, \eta^2 = .05)$. This increase appeared to be localized to the later minutes of DAA:Double task, but it must be noted that this effect was statistically very small. This modest increase did not push PICs anywhere near an overload workload state, as is visible in Figure \ref{Fig:OverallWorkload-Trial}.  A similar pattern was observed when the number of active UAS increased, where a higher number of active UAS produced a reliable, but statistically small, increase in the estimated overall workload $(F(531,2655) = 1.25, p < .001, \eta^2 = .02)$, which again was constrained mostly to later minutes of the DAA:Double task.  

\paragraph{Weather}
The 10:Low and 24:High trial's Weather overall estimated mean workload results tended to be slightly higher than the corresponding DAA:Double task results, while the 24:Low Weather results were generally equivalent to that trial's DAA:Double results. However, all Weather tasks' results were lower than the DAA:Single tasks' workload results. The change in overall estimated workload based on the number of nests was not statistically reliable overall,
but there did appear to be a significant change in estimated overall workload over minutes and seconds during the Weather event based on increasing the number of nests $(F(531,2124) = 1.15, p = .02, \eta^2 = .05)$. This change in overall workload was focused in the later minutes of the task and was statistically reliable with a very small effect, and no PICs' results were near an overload state.  Relative to the number of active UAS, while Weather did produce a small, but reliable, overall increase in workload $(F(1,5) = 8.49, p = .03, \eta^2 = .05)$, increasing the number of active UAS did not impact the degree of this change. 

\subsubsection{Subjective In Situ Workload}

The PICs provided ratings for each workload component at 3m and 6m during each task. While there are well documented limitations of subjective responses, the in situ workload probes provide insights into the PICs' perceived workload. The presented results are normalized (i.e.,\ overall workload raw value/[49 = 7 point Likert scale x 7 workload components], and component workload raw value/7 point Likert scale) to facilitate comparison to the objective workload estimates. There are a total of 36 in situ responses (i.e.,\ 2 prompts x 6 tasks x 3 PICs) for the 10:Low and 24:Low conditions, and 72 for the 24:High condition (i.e.,\ 2 prompts x 6 tasks x 6 PICs). This number of data points does not provide the necessary statistical power; thus, only descriptive statistics are provided for these results.  

The in situ overall workload results by trial were in the very low-normal workload range (i.e.,\ close to 20) for the 10:Low and 24:Low trials, with the 24:High trial results often in the underload range (i.e.,\ $< 20$), as shown in Table \ref{Table:InSituOverallWorkload}. These results have a broader range than the objective workload estimates (provided in Table \ref{Table:OverallWorkload}, but are within one standard deviation. The results also show that the PICs subjectively rated their workload highest for the first trial they completed, either 10:Low or 24:Low. While the differences between these X:Low trials and 24:High was not very large for the objective workload estimates, the subjective responses indicated lower workload for the 24:High trial. 

\begin{table}[hbt]
\centering
\caption{The normalized in situ overall workload results descriptive statistics by trial. 
} 
\label{Table:InSituOverallWorkload}
\begin{tabular}{|c|c|c|}
\hline
\textbf{10:Low} & \textbf{24:Low} & \textbf{24:High} \\ \hline
 23.98 (7.27) & 20.8 (2.73) & 18.43 (6.18) \\ \hline
\end{tabular}
\end{table}

The mean in situ overall workload tended to be highest during the Nominal \#1 task for the 10:Low and 24:Low trials, with subsequent slight decreases until the Weather task, as shown in Table \ref{Table:InSituOverallWorkloadSummary}. This result matches the pattern found for the objective workload estimates (see Table \ref{Table:WorkloadTasks}). The 24:High condition's in situ results were flat across the first five tasks, with a very slight increase during the Weather task. While the X:Low trial's in situ overall workload is mostly higher than the 24:High trial, all values are within one standard deviation and are not generally different. 

\begin{table}[hbt]
\centering
\caption{The normalized in situ overall workload descriptive statistics by trial. } 
\label{Table:InSituOverallWorkloadSummary}
\resizebox{\columnwidth}{!}{
\begin{tabular}{|l|c|c|c|c|c|c|}
\hline
\multirow{2}{*}{\textbf{Trial}} & \textbf{Nominal} & \textbf{DAA:} & \textbf{Nominal} & \textbf{DAA:}  & \textbf{Nominal} & \multirow{2}{*}{\textbf{Weather}} \\ 
 & \textbf{\#1} & \textbf{Single} & \textbf{\#2} & \textbf{Double}  & \textbf{\#3} & \\ \hline
\hline
\textbf{10:Low} & 28.92 (4.61) & 25.86 (6.96) & 22.45 (8.90) & 22.45 (8.24) & 21.76 (7.78) & 22.80 (7.12) \\
\hline
\textbf{24:Low} & 27.20 (3.33) & 23.81 (3.14) & 20.06 (4.84) & 20.06 (2.35) & 17.69 (1.18) & 18.73 (1.61) \\
\hline
\textbf{24:High} & 18.37 (6.18) & 18.53 (6.08) & 18.37 (6.24) & 18.37 (6.29) & 18.02 (6.39) & 19.04 (5.94) \\
\hline
\end{tabular}
}
\end{table}

The 10:Low and 24:Low Nominal tasks showed a clear trend with the in situ overall workload ratings decreasing with each successive Nominal task (\#1 - \#2 - \#3), with larger decreases between the 24:Low Nominal tasks. The 24:High trial had the lowest Nominal \#1 and \#2 task overall workload means, with no meaningful decline across the Nominal tasks. 

The highest in situ overall mean workload during the unexpected events occurred primarily for the DAA:Single task for the 10:Low and 24:Low trials, and for the 24:High trial Weather event. These results demonstrate a similar pattern to the objective overall workload estimates. The DAA:Double task's in situ overall mean workload was effectively the same as the Nominal tasks and the Weather event. 

The subjective workload responses generally support the expectations that the visual and cognitive workload components were loaded the most for PICs. There was virtually no difference between and withing the remaining component results, as shown in Table \ref{Table:InSituWorkloadComponents}, given the large standard deviations. However, it is noted that the subjective responses did decrease from the PICs' first trial (i.e.,\ 10:Low or 24:Low) to their 24:High trial for the cognitive, visual, fine motor and tactile workload components. These decreases were represented in the in situ overall workload ratings. 

\begin{table}[hbt]
\caption{The normalized in situ workload descriptive statistics by workload component. } 
\label{Table:InSituWorkloadComponents}
\resizebox{\columnwidth}{!}{
\begin{tabular}{|l|c|c|c|c|c|c|c|}
\hline
\multirow{2}{*}{\textbf{Trial}} & \multirow{2}{*}{\textbf{Cognitive}} & \multirow{2}{*}{\textbf{Visual}} & \multirow{2}{*}{\textbf{Speech}} & \multirow{2}{*}{\textbf{Auditory}}  & \textbf{Gross} & \textbf{Fine} & \multirow{2}{*}{\textbf{Tactile}} \\ 
 & & & & & \textbf{Motor} & \textbf{Motor} &\\ \hline
10:Low  &  37.71 (10.86) & 33.71 (8.71) & 19.0 (8.29)  & 19.86 (8.29) &   19.43 (8.29) & 19.0 (8.29) & 19.0 (8.29)     \\ \hline
24:Low  & 30.57 (11.57) & 34.14 (6.57) & 15.86 (2.71) & 14.71 (0.71) & 14.29 (0.0) & 19.0 (4.0) & 17.0 (4.14) \\ \hline
24:High & 23.57 (11.71) & 22.0 (7.57) & 16.71 (5.86) & 16.71 (5.86) & 16.71 (5.86) & 16.71 (5.86) & 16.71 (5.86)  \\ \hline
\end{tabular}
}
\end{table}

\subsection{User Interactions with Displays/Materials}

A comprehensive view of PIC performance requires understanding their interactions with the displays. The video recordings from both the eye tracker's forward facing camera, available camcorder footage (only three PICs), and the ADS-B screen captures were analyzed to document how PICs interacted with and queried the various displays/materials. Each interaction, usually indicated by a computer mouse action, was identified and coded based on which of the four displays were interfaced (i.e.,\ ADS-B, Wing Interface, METAR, Chat), or the Offscreen materials (e.g.,\ PIC Reference Sheet and Procedures Manual).  The broad nature or type of action (i.e.,\ click on menu, zoom in) was also recorded.  This interaction video coding was completed for all tasks across both trials for all six PICs.

The overall number of interactions per trial was highly variable across PICs and trials (minimum 83, maximum 548), and averaged 230.66 interactions (std = 132.17) per trial.  Independent of task or trial, the majority of interactions involved one of three areas, as shown in Table \ref{Table:InterCountOverall}: the ADS-B display (72\% of total interactions), Wing Interface (17\% of total interactions), or Offscreen materials (12\% of total interactions).  The METAR and Chat were very rarely interacted with ($<1\%$ of total interactions combined).  

\begin{table}[!hbt]
\centering
\caption{The total interaction counts by trial and display/resource. } 
\label{Table:InterCountOverall}
\resizebox{\columnwidth}{!}{
\begin{tabular}{|c|c|c|c|c|c|c|}
\hline
 \multirow{2}{*}{\textbf{Trial}} & \multirow{2}{*}{\textbf{METAR}} & \textbf{Wing Delivery} & \textbf{ADS-B} & \multirow{2}{*}{\textbf{Chat}} & \multirow{2}{*}{\textbf{Offscreen}}  & \multirow{2}{*}{\textbf{Total}}  \\ 
 & & \textbf{Interface} & \textbf{Display} & & & \\ \hline
\textbf{10:Low} &  0 & 52	& 336	& 0	&  156 & 544\\ \hline
\textbf{24:Low} & 0	& 161	& 737	& 0	&  133 & 1,031\\ \hline
\textbf{24:High} & 1 & 257	& 938	& 0	&  38 & 1,234\\ \hline
\textbf{Total} & 1	& 470	& 2,011	& 0	&  327& 2,809\\ \hline
\end{tabular}
}
\end{table}

The majority of ADS-B interactions, 78\% involved PICs querying the interface for details about crewed Aircraft, as shown in Table \ref{Table:ADSBInter}.  The next most common ADS-B interaction was Zooming the field of view in or out (13\%), while the remaining actions (i.e.,\ Mouse as Focus, adjusting Settings, or Panning) were more infrequent ($<4\%$ each).  When considering interactions with the Wing Interface, as shown in Table \ref{Table:WingInter}, 79\% of the interactions involved PICs accessing the Menu (i.e., aircraft cards, fleet view) that contained information about Wing aircraft 
Approximately 11\% of the Wing interface interactions involved adjusting Settings for the display of information, and the remaining activities were infrequent.  Finally, the interaction with Offscreen materials seemed to be split nearly evenly between reviewing Paper Materials that contained mission parameters (58\%) and asking Verbal Questions to clarify information (42\%).

\begin{table}[!hbt]
\centering
\caption{The ADS-B display total interaction counts by trial, action, and task. } 
\label{Table:ADSBInter}
\begin{tabular}{|c|c|c|c|c|c|c|}
\hline
\multirow{2}{*}{\textbf{Trial}} & {\textbf{Crewed Aircraft}} & \textbf{Mouse} & \multirow{2}{*}{\textbf{Settings}} & {\textbf{Zoom}} & \multirow{2}{*}{\textbf{Pan}} & \multirow{2}{*}{\textbf{Total}}  \\ 
 & \textbf{Information} & \textbf{as Focus} &  & \textbf{In/Out} & & \\ \hline
 \multicolumn{7}{|c|}{\textbf{Nominal \#1}} \\ \hline
\textbf{10:Low} &  44	& 3	& 0	& 0	& 0 & 47\\ \hline
\textbf{24:Low} &  61	& 3	& 11 & 36	& 4 & 115 \\ \hline
\textbf{24:High} & 87  & 13	& 10	& 24	& 3 & 137\\ \hline
 \multicolumn{7}{|c|}{\textbf{DAA: Single}} \\ \hline
\textbf{10:Low} & 61	& 2	& 0	& 0	& 0 & 63\\ \hline
\textbf{24:Low} & 102	& 5	& 13 &	30	& 8 & 158\\ \hline
\textbf{24:High} & 187	& 19	& 0	& 23	& 8 & 237 \\ \hline
 \multicolumn{7}{|c|}{\textbf{Nominal \#2}} \\ \hline
\textbf{10:Low} &  48	& 10	& 0	& 0	& 0 & 58\\ \hline
\textbf{24:Low} & 50	& 0	& 3	& 10	& 12 & 75\\ \hline
\textbf{24:High} & 97	& 4	& 0	& 28	& 0 & 129 \\ \hline
 \multicolumn{7}{|c|}{\textbf{DAA: Double}} \\ \hline
\textbf{10:Low} & 42	& 2	& 0	& 0	& 0 & 44\\ \hline
\textbf{24:Low} &  86	& 4	& 0	& 12	& 11 & 113\\ \hline
\textbf{24:High} & 105	& 3	& 0	& 17	& 1 & 126\\ \hline
 \multicolumn{7}{|c|}{\textbf{Nominal \#3}} \\ \hline
\textbf{10:Low} & 72	& 0	& 0	& 0	& 0 & 72\\ \hline
\textbf{24:Low} &  106	& 0	& 4	& 29	& 11 & 150\\ \hline
\textbf{24:High} & 152	& 2	& 0	& 20	& 0 & 174\\ \hline
 \multicolumn{7}{|c|}{\textbf{Weather}} \\ \hline
\textbf{10:Low} & 51	& 0	& 0	& 0	& 0 & 51 \\ \hline
\textbf{24:Low} & 99	& 6	& 2	& 15	& 4 & 126\\ \hline
\textbf{24:High} & 124	& 0	& 0	& 11	& 0 & 135\\ \hline
\multicolumn{7}{|c|}{\textbf{Total Overall}} \\ \hline
\textbf{Total} & 1,574	& 76	& 43	& 255	& 62& 2,010\\ \hline
\end{tabular}
\end{table}

\begin{table}[!hbt]
\centering
\caption{The total interaction counts with subsections of the Wing Interface by trial, action, and task. } 
\label{Table:WingInter}
\begin{tabular}{|c|c|c|c|c|c|c|}
\hline
\multirow{2}{*}{\textbf{Trial}} & {\textbf{Wing Aircraft}} & \textbf{Mouse} & \textbf{Zoom} & \multirow{2}{*}{\textbf{Settings}} & \multirow{2}{*}{\textbf{Pan}} & \multirow{2}{*}{\textbf{Total}}  \\ 
 & \textbf{Menu}  & \textbf{as Focus} & \textbf{In/Out} &  & & \\ \hline
 \multicolumn{7}{|c|}{\textbf{Nominal \#1}} \\ \hline
\textbf{10:Low} &  21	& 8	& 0	& 1	& 0 & 30\\ \hline
\textbf{24:Low} &  14	& 0	& 2 & 4	& 6 & 26 \\ \hline
\textbf{24:High} & 9  & 0	& 2	& 6	& 9 & 26\\ \hline
 \multicolumn{7}{|c|}{\textbf{DAA: Single}} \\ \hline
\textbf{10:Low} & 6	& 0	& 0	& 6	& 0 & 12\\ \hline
\textbf{24:Low} & 16	& 0	& 0 &	3	& 0 & 19\\ \hline
\textbf{24:High} & 53	& 0	& 0	& 4	& 2 & 59 \\ \hline
 \multicolumn{7}{|c|}{\textbf{Nominal \#2}} \\ \hline
\textbf{10:Low} &  0	& 0	& 0	& 0	& 0 & 0\\ \hline
\textbf{24:Low} & 48	& 0	& 4	& 1	& 1 & 54\\ \hline
\textbf{24:High} & 33	& 0	& 0	& 0	& 0 & 33 \\ \hline
 \multicolumn{7}{|c|}{\textbf{DAA: Double}} \\ \hline
\textbf{10:Low} & 0	& 0	& 0	& 0	& 0 & 0\\ \hline
\textbf{24:Low} &  22	& 0	& 0	& 2	& 0 & 24\\ \hline
\textbf{24:High} & 0	& 0	& 0	& 0	& 0 & 0\\ \hline
 \multicolumn{7}{|c|}{\textbf{Nominal \#3}} \\ \hline
\textbf{10:Low} & 0	& 0	& 0	& 0	& 0 & 0\\ \hline
\textbf{24:Low} &  32	& 0	& 0	& 0	& 0 & 32\\ \hline
\textbf{24:High} & 19	& 0	& 0	& 2	& 0 & 21\\ \hline
 \multicolumn{7}{|c|}{\textbf{Weather}} \\ \hline
\textbf{10:Low} & 6	& 0	& 0	& 4	& 0 & 10 \\ \hline
\textbf{24:Low} & 1	& 0	& 0	& 5	& 0 & 6\\ \hline
\textbf{24:High} & 56	& 5	& 2	& 8	& 2 & 73\\ \hline
\multicolumn{7}{|c|}{\textbf{Total Overall}} \\ \hline
\textbf{Total} & 336	& 13	& 10	& 46	& 20& 425\\ \hline
\end{tabular}
\end{table}

\begin{table}[!hbt]
\centering
\caption{The total interaction counts with Offscreen Materials by trial, action, and task. } 
\label{Table:OffscreenInter}
\begin{tabular}{|c|c|c|c|}
\hline
\multirow{2}{*}{\textbf{Trial}} & {\textbf{Paper}} & \textbf{Verbal} &  \multirow{2}{*}{\textbf{Total}}  \\ 
 & \textbf{Manuals} & \textbf{Question} & \\ \hline
 \multicolumn{4}{|c|}{\textbf{Nominal \#1}} \\ \hline
\textbf{10:Low} &  12	& 4	& 16	\\ \hline
\textbf{24:Low} &  20	&27	& 47  \\ \hline
\textbf{24:High} & 10  & 12	& 22	\\ \hline
 \multicolumn{4}{|c|}{\textbf{DAA: Single}} \\ \hline
\textbf{10:Low} & 17	& 31	& 48	\\ \hline
\textbf{24:Low} & 24	& 17	& 41 \\ \hline
\textbf{24:High} & 4	& 0	& 4	 \\ \hline
 \multicolumn{4}{|c|}{\textbf{Nominal \#2}} \\ \hline
\textbf{10:Low} &  24	& 15	& 39	\\ \hline
\textbf{24:Low} & 12	& 10	& 22	\\ \hline
\textbf{24:High} & 4	& 0	& 4	 \\ \hline
 \multicolumn{4}{|c|}{\textbf{DAA: Double}} \\ \hline
\textbf{10:Low} & 25	& 3	& 28	\\ \hline
\textbf{24:Low} &  6	& 9	& 15	\\ \hline
\textbf{24:High} & 4	& 0	& 4	\\ \hline
 \multicolumn{4}{|c|}{\textbf{Nominal \#3}} \\ \hline
\textbf{10:Low} & 14	& 0	& 14	\\ \hline
\textbf{24:Low} &  0	& 0	& 0	\\ \hline
\textbf{24:High} & 0	& 0	& 0	\\ \hline
 \multicolumn{4}{|c|}{\textbf{Weather}} \\ \hline
\textbf{10:Low} & 0&11	& 11	 \\ \hline
\textbf{24:Low} & 9	& 0	& 9	\\ \hline
\textbf{24:High} & 4	& 0	& 4	\\ \hline
\multicolumn{4}{|c|}{\textbf{Total Overall}} \\ \hline
\textbf{Total} & 189	& 139	& 328	\\ \hline
\end{tabular}
\end{table}

\subsubsection{Number of Nests}
Increasing the number of nests did appear to impact the overall number of user interactions.  The total number of interactions in the 24:Low trial (1034 interactions) was roughly 90\% higher than in the 10:Low trial (544 interactions).  While the number of overall interactions was higher in the 24:Low trial, proportionally users interacted in X:Low conditions in very similar ways.  The 10:Low condition PICs interacted with the ADS-B display 62\% of the time, whereas the 24:Low PICs interacted with the ADS-B display 71\% of the time.  Interactions with the ADS-B display during the 10:Low condition focused almost exclusively on obtaining crewed aircraft information (95\%), which was less pronounced for 24:Low (68\%) that also contained more Zooming in and out (18\%).  The proportion of overall interactions with the Wing Interface was also very similar across the 10:Low (10\%) and 24:Low (16\%) trials.  Information gathering about the details of active UAS was the primary activity in both the 10:Low (63\%) and 24:Low (83\%) conditions. Adjusting system settings was the second most likely interaction with the Wing user interface for both the 10:Low (21\%) and 24:Low (9\%) PICs.  However, the 10:Low PICs interacted with Offscreen materials (29\%) at a much higher rate than the 24:Low condition PICs (13\%), an increase of roughly 123\%.  Despite this overall difference in proportion of Offscreen interaction across the different nest conditions (as shown in Table \ref{Table:OffscreenInter}), the type of interactions appear consistent across the groups as the 10:Low (59\%) and 24:Low (53\%) groups referred to the Paper Documentation nearly equivalently.  There were no differences across the X:Low conditions for the remaining systems (i.e., METAR, Chat).

The interactions during the Nominal tasks resulted in more user interactions during the 24:Low condition (524 interactions) vs.\ the 10:Low condition (277 interactions); a 90\% increase.  However, there was almost no difference in the proportional interactions with the ADS-B display between the 10:Low (64\%) and 24:Low (65\%) conditions. Specific to the ADS-B display, the 10:Low PICs appeared to focus exclusively on crewed aircraft information (93\%), which was less pronounced for the 24:Low PICS (64\%).  There did appear to be more Zooming in and out on the ADS-B interface for the 24:Low PICs (22\%) as compared to the 10:Low PICs (0\%).  The 24:Low PICs did tend to interact with the Wing Interface slightly more (21\%) than the 10:Low PICs (11\%).  The 24:Low PICs also viewed active Wing aircraft information more frequently (84\%) than the 10:Low (70\%) PICs. This pattern was reversed for the Offscreen materials with the 10:Low PICs interacting with these materials at a higher rate (25\%) than the 24:Low group (13\%).  PICs in the 10:Low condition were much more likely to review the Paper documentation (72\%) than those in the 24:Low (46\%) group, but the 24:Low PICs asked more Verbal questions (54\%).  There were no substantive interaction differences across the X:Low conditions for the other displays.

\subsubsection{Number of Active UAS}
Increasing the number of active UAS from the X:Low to the 24:High trials did not produce a corresponding increase in user interactions, independent of task.  The X:Low trial PICs averaged 263 total interactions (std = 163.43), but in the 24:High trials only averaged 205.67 (std = 93.12) interactions; a drop of nearly 22\%.  Despite fewer interactions overall during the 24:High trial, proportionally the amount of overall interactions with the ADS-B display were roughly equivalent across the X:Low (68\%) and 24:High (79\%) trials.  Proportionally the X:Low and 24:High trials were equivalent in their type of interactions with the ADS-B display,  as shown in Table \ref{Table:ADSBInter}.  Interactions with the Wing interface were also similar in rate across the X:Low (14\%) and 24:High (21\%) trials, and the nature of the type of interaction was nearly identical across these trials.  There was a slight difference in how much interaction with Offscreen materials, with PICs in the X:Low trials engaging with these materials slightly more often (18\%) than during the 24:High (3\%) trials.  This outcome is likely due to the relative unfamiliarity of the evaluation parameters during the X:Low trials, which caused PICs to review the written Reference Sheet (56\%) and ask Verbal questions (44\%) more frequently.  The 24:High trial's Offscreen interactions focused more on the Paper reference materials (68\%) and asked proportionally fewer verbal questions (32\%).  

The analysis of the Nominal tasks found the interaction pattern was largely consistent with the overall interaction results.  The ADS-B display was interacted with the most, although slightly more frequently in the 24:High trial (440 interactions, 80\% of the task's total interactions) than during the X:Low trials (517 interactions, 65\% of the task's total interactions).  
The type of interaction with the ADS-B display was nearly identical across the X:Low and 24:High trials, with a primary focus on accessing aircraft information.  The Wing interface was also proportionally interacted with at similar rates across the X:Low (142 interactions, 18\%) and 24:High (80 interactions, 15\%) trials, and both trials produced nearly identical patterns of interaction, again with aircraft information as the largest and primary focus.  Offscreen materials were interacted with more during the X:Low trials (138 interactions, 17\%) than in the 24:High trial (26 interactions, 4\%).  Both the X:Low and 24:High trials reviewed Offscreen materials in similar ways and split interactions evenly between paper materials and verbal questions.

\subsubsection{Unexpected Events}
The introduction of the unexpected events was anticipated to change how PICs interacted with the displays/materials at hand.  Specifically, these interactions were expected to reflect more information gathering activities as PICs were making decisions regarding how best to handle each unexpected event.

\paragraph{Detect and Avoid: Single Crewed Aircraft}
The injection of the DAA:Single event did appear to increase the number of interactions.  Independent of trial, while the average overall number of interactions in the Nominal tasks was 37.44 (std = 26.27), the DAA:Single event's number increased to 53.42 (std = 33.35), a roughly 43\% increase.  Interactions with the ADS-B display reflected 71\% of these interactions, whereas interactions with the Wing interface encompassed 14\%, and Offscreen materials 15\%.  PICs did not interact with the other displays at all during the DAA:Single event.  The DAA:Single trial interactions were virtually identical to the Nominal task proportional averages (ADS-B: 65\%, Wing interface: 16\%, Offscreen materials: 19\%), suggesting that while more overall interactions occurred during the DAA:Single task, proportionally the PICs were interacting in similar ways.  This outcome appears to represent the PICs simply gathering more information, and not necessarily different information gathering behavior.

Increasing the number of nests impacted the amount of interactions during the DAA:Single task. The 10:Low condition averaged 41 interactions (std = 19.70), while the 24:Low condition averaged 72.67 (std = 43.06), an increase of 77\%.  Proportionally, the number of nests also increased how PICs interacted with the ADS-B display.  51\% of the 10:Low PIC interactions were with the ADS-B display; however, 72\% of the 24:Low interactions were with the ADS-B, an increase of 41\%. While interacting with the ADS-B display, the 10:Low PICs focused nearly exclusively on crewed aircraft information (97\%), which was much higher than the 24:Low PICs (65\%), whose interaction pattern was consistent with the Nominal task interactions.  There was virtually no difference in the overall proportion of PICs interactions with the Wing interface in the 10:Low (10\%) and 24:Low (9\%) conditions, although the 24:Low group was much more concerned with Wing aircraft information (84\%) vs. the 10:Low group (50\%). PICs in the 10:Low condition interacted with the Offscreen materials (39\%) much more than individuals in the 24:Low condition (19\%).  PICs in the 10:Low condition were less likely to review the Paper guidelines (35\%) than the 24:Low group (59\%).

The increase in number of active UAS did not increase the amount of interaction during a DAA:Single task.  The X:Low trials averaged 56.83 interactions (std = 34.61), while the 24:High trial averaged 50 (std = 34.96), both were similar to the Nominal tasks.  There did appear to be a slight proportional increase in the 24:High trial's interactions with the ADS-B display (79\%) vs. the X:Low condition (65\%), although the interaction types were virtually identical proportionally across possible interactions, with a consistent focus on crewed aircraft information (74\% and 78\%, respectively).  
Similarly, the 24:High trial produced more interactions with the Wing interface (20\%) than the X:Low trials (9\%).  The 24:High trial's interactions were more focused on Wing aircraft information (90\%) than the X:Low trials (71\%). 
Finally, the X:Low trials resulted in more Offscreen interactions (26\%) than the 24:High trial (1\%); however, the nature of these interactions was equivalent across trials with a nearly equal split between reviewing paper materials and verbal questions.

\paragraph{Detect and Avoid: Double Crewed Aircraft}
The onset of a DAA:Double event did not increase the PIC interactions.  The PICs overall produced, on average, 29.5 interactions (std = 16.87), which was approximately 21\% less than the Nominal task average, and roughly 45\% less than the number of overall interactions that occurred during the DAA:Single task.  Nearly 80\% of all interactions were with the ADS-B display, while only 7\% involved the Wing interface, and 13\% were Offscreen.  The PICs did not interact with the other systems at all.

Increasing the number of nests did lead to increased interactions during the DAA:Double task.  The 10:Low condition averaged 24 interactions (std = 9), while the 24:Low condition averaged over twice that with 50.67 interactions (std = 17.62).  Proportionally,  the 24:Low PICs interacted with the ADS-B display at a slightly higher rate (74\%) than the 10:Low PICs (61\%).  The 10:Low PICs expended considerably more effort gathering information regarding crewed aircraft (95\%) vs. the 24:Low PICs (76\%). The 24:Low PICs interacted with the Wing interface at a higher rate (16\%) than the 10:Low condition, who did not interact with it at all (0\%).  The interaction with the Wing interface during the 24:Low condition was almost exclusively focused on Wing aircraft (92\%).  Related to the Offscreen materials, the 10:Low PICs engaged with these materials at a much higher level (39\%) than the 24:Low PICs (10\%).  The 10:Low PIC interactions with these materials was focused on Paper guidelines at a very high rate (90\%), much more so than the 24:Low group (40\%).

The increase in active UAS resulted in fewer overall interactions during the DAA:Double task.  The 24:High trial averaged 21.67 interactions (std = 10.42), which was approximately 42\% lower than the X:Low trials (mean = 37.33, std = 19.23).  The 24:High PICS focused their interactions predominately on the ADS-B display (97\%), with limited interactions with the Offscreen materials (3\%).  Interactions with the ADS-B display focused on crewed aircraft 83\% of the time, while Zooming was the second most likely action (13\%).  Conversely, the X:Low DAA:Double tasks were more consistent with overall and Nominal task interaction patterns.  The X:Low trials focused 70\% of their interactions on the ADS-B display, 11\% on the Wing interface, and 19\% on Offscreen materials.  Interactions with the ADS-B display for the X:Low trials also often focused on crewed aircraft information (82\%) at a rate consistent with the 24:High trial.  92\% of interactions with the Wing interface during the X:Low DAA:Double task were related to the Wing aircraft information. Both the X:Low and 24:High trials exhibited a preference for reviewing printed materials instead of asking verbal questions.  There was no interaction with the other systems in the X:Low trial.

\paragraph{Weather}
The Weather unexpected event did not generally increase the levels of interaction.  The PICs on average produced 35.42 interactions (std = 24.53), which was nearly identical to Nominal tasks and the DAA:Double task, but still 34\% lower than the DAA:Single task.  Proportionally, the interaction patterns were similar to Nominal tasks, where PICs interacted most frequently with the ADS-B display (73\%), and second-most often with the Wing interface (21\%).  PICs interacted with the Offscreen materials (6\%) at a slightly lower rate than during the Nominal tasks, likely due to the Weather task occurring last in each trial, after they had familiarized themselves with the operational procedures.

The increase in number of nests did increase the amount of interaction during the Weather task.  The 10:Low condition averaged 24 interactions (std = 15.39), which was slightly less than half of the 47 interactions (std = 39.89) observed on average during the 24:Low condition.  The 24:Low condition also proportionally resulted in higher rate (89\%) of ADS-B display interactions as compared to the 10:Low condition (71\%).  However, the 10:Low PICs' interaction with the ADS-B display focused entirely on crewed aircraft information (100\%), while the 24:Low group focused on this type of interaction at more normal rate (79\%).  The 24:Low group was more likely to Zoom the ADS-B display in or out  (12\%) than the 10:Low PICs during the Weather task.  The X:Low conditions were reversed for the Wing interface and Offscreen materials, as the 10:Low PICs focused on these resources (14\%, 15\%, respectively) at slightly higher rate than the 24:Low condition (4\%, 6\%, respectively).  While there were very few interactions with the Wing interface, the 10:Low group more frequently gathered crewed aircraft information (60\%) than the 24:Low PICs (17\%).  The 24:Low PICs were also much more likely to interact with paper Offscreen materials (100\%), while the 10:Low PICs chose to exclusively ask verbal questions during the weather task (100\%).  Neither condition interacted with the other available systems.

There was virtually no increase in the amount of interaction due to increasing the number of active UAS during the Weather task.  The X:Low trials averaged 35.5 interactions (std = 29.83), while the 24:High trials averaged 35.33 interactions (std = 20.81).  Despite the virtually equivalent overall number of interactions, there were slight differences with what the PICs interacted with.  The X:Low trials produced a proportionally higher number of interactions with the ADS-B display (83\%) than the 24:High trial (64\%).  However, the ADS-B display interactions for both the X:Low (85\%) and 24:High (92\%) trials were equally likely to be focused on crewed aircraft information.  Similarly, the X:Low trials produced more interactions with these Offscreen materials (9\%) than the 24:High trials (2\%).  Interestingly, the 24:High trial produced over a 4x higher rate of interaction with the Wing interface (34\%) than the X:Low trials (8\%).  Specific to the Wing interface, the 24:High PICs focused on Wing aircraft information at roughly normal levels (77\%), while the X:Low PICS did so only 44\% of the time.  No other interactions were observed with the remaining systems.

\subsection{Situation Awareness}
The PICs have a limited number of decisions for which they can take action on in the system since order assignments to the UAS and the actual order deliveries are completed autonomously. Essentially, the PICs can either pause the system to ensure no additional UAS depart for a delivery or can issue a land now command. Issuing a land now command occurs very rarely, and was excluded from this evaluation in order to isolate monitoring and intial action response and reduce variability. The locus of attention results (see Section \ref{Sec:Focus}) demonstrated that the PICs primarily focus on the ADS-B display, with some checks on the Weather, to determine if operations needed to be paused. 

Overall, the situation awareness (SA) probe response correctness (percent correct) was very high. There were eighteen SA probes during each trial. Across all trials, the probe responses had $> 92\%$ accuracy, as shown in Table \ref{Table:SAProbeOverallTrial}. The highest number of correct responses across the trials occurred for the SA$_{2}$ probe, shown in Table \ref{Table:SAProbeOverallTrialProbe}. All trials had instances in which PICs answered that they did not intend to change the system status (i.e.,\ SA$_{3:4:30}$) or operability (i.e.,\ SA$_{3:8:00})$ in the next few minutes, resulting in a modest reduction in the accuracy percentages. 

\begin{table}[!hbt]
\centering
\caption{The SA probe correctness percentages.}
\label{Table:SAProbeOverall}
\begin{subtable}[t]{1\textwidth}
\centering
\begin{tabular}[t]{|c|c|c|}
\hline
\textbf{10:Low} & \textbf{24:Low} & \textbf{24:High} \\ \hline
 92.59\% & 92.59\% & 96.30\% \\ \hline 
\end{tabular}
\caption{Percentages by trial.}
\label{Table:SAProbeOverallTrial}
\end{subtable}
\begin{subtable}[t]{1\textwidth}
\centering
\begin{tabular}[t]{|c|c|c|c|}
\hline
 & \textbf{SA$_{2}$} & \textbf{SA$_{3-4:30}$} & \textbf{SA$_{3-8:00}$} \\ \hline
\textbf{10:Low} & 94.44\% & 88.89\% & 94.44\% \\
\textbf{24:Low} & 100\% & 88.89\% & 88.89\% \\
\textbf{24:High} & 97.22\% & 97.22\% & 94.44\% \\ \hline 
\end{tabular}
\caption{Percentages by trial and SA Probe.}
\label{Table:SAProbeOverallTrialProbe}
\end{subtable}
\end{table}

The by task analysis found 100\% correct responses to all three SA probes for each of the three Nominal tasks and the DAA:Single task. The same PIC incorrectly answered the SA$_{2}$ probes during the DAA:Double task for both trials (i.e., two incorrect answers) when the encounter vehicle was present on the ADS-B display. The incorrect SA$_{3-8:00}$ response was a case where the first of the two crewed aircraft encounters was present on the ADS-B display and noticed by the PIC. At the time, the deliveries were not paused, and the PIC responded ``maybe'' in response to an expected change in the system operability in the next two minutes. This response may be perceived as partially correct, but since the DAA:Double was designed to ensure PIC's did not pause the system, this response was coded as incorrect. 

The majority of the incorrect responses occurred during the Weather task for which the SA$_{3}$ probes required that the PIC must have noticed the onset of adverse weather and paused the system's operation prior to responding to the SA$_{3-X}$ probes. If the system operations were not paused and the PIC responded ``No'', this response was incorrect, since the PIC had not noticed or reacted to the weather change. The Weather was consistently updated at 52 minutes and 20 seconds into the trial (or 2 minutes and 20 seconds into the task), which is $> 2$ minutes prior to the onset of the SA$_{3-4:30}$ probe. Four PICs had incorrect responses for the SA$_{3-4:30}$ probe in their respective X:Low conditions, two in each. One 10:Low condition PIC noticed the Weather change and paused the system prior to responding correctly to the SA$_{3-8:00}$ probe, while the others responded incorrectly. Additionally, one of these PICs incorrectly responded to the SA$_{3:8:00}$ probe during the 24:High condition. 

Effectively, the SA probe response accuracies were very good. There were no differences by the Number of Nests, and the percent correct improved with more active UAS (i.e.,\ 24:High) for five PICs. This marked improvement is likely due to learning effects that occurred during the X:Low trials related to the Weather task. Three incorrect responses occurred during the DAA:Double unexpected event task; however, the one PIC's incorrect responses to the SA$_{2}$ probe in both trials is indicative of a different issue.

\section{Discussion} \label{sec-Discussion}
The PICs focused their attention predominately on the ADS-B display followed by the Wing user interface. This prioritization of the PICs' locus of attention was the same across trials, tasks, number of nests and number of active UAS. The unexpected events did result in increased attention placed on the most relevant displays, the ADS-B for the DAA:X tasks and the METAR during the Weather task. An analysis of the ADS-B displays AOIs found that there was generally no substantial change in the locus of attention across the trials, unless a DAA:X encounter occurred in spatially different AOIs. The overall locus of attention results demonstrate that the PICs were engaged with each task during each trial in a manner consistent with their job responsibilities. 

The workload experienced by the PICs across conditions presented a fairly consistent picture in both the objective and subjective metrics. Specifically, estimated overall workload remained well within normal workload range, with a few instances falling into the underload range. At no time during any of the conditions did the PIC estimated overall workload go into the overload range. Under nominal conditions, increasing the number of nests, or the number of active UAS did not produce any detectable increase in estimated overall workload. The DAA:Single task resulted in slight higher, but statistically unreliable estimated overall workload, suggesting that number of nests and number of UAS did not increase estimated overall workload even when experiencing the crewed aircraft encounter.  The number of nests did modestly increase estimated overall workload in the DAA:Double cases, as did number of active UAS for the DAA:Double task, however, it must be emphasized that these increases were statistically small and estimated overall workload remained well within nominal limits. While it did seem that more nests may slightly impact the estimated overall workload during a Weather event, the number of active UAS did not seem to influence estimated overall workload during the Weather conditions.
These small fluctuations in estimated overall workload were exactly that, and did not at all suggest that PICs experienced levels of workload that fall into the overload range. As expected, the cognitive and visual workload components resulted in the largest contributions to estimated overall workload. 

Overall, the normalized subjective workload results show the PICs perceived their workload was underloaded across all conditions. The higher perceived cognitive and visual workload results indicated that PICs felt these components contributed the most to their overall workload. These results agree with the estimated workload and IMPRINT Pro model results. 

The objective workload estimates and perceived workload results were expected to be substantially lower than the IMPRINT Pro model predicted results. The model's substantially higher predicted workload components and overall workload were due to limitations of the modeling tool and the lack of accurate workload models for use cases similar to the $1:M$ delivery use case.  

The PICs generally maintained very good overall situation awareness throughout the trials. Situation awareness fell during the DAA:Double and Weather tasks, however the Weather results were influenced by the pre-session briefings. Overall the results corroborate the attention and workload results that generally indicate the number of nests, active UAS and the unexpected events did not overly impact the PICs' performance.

PICs also interacted with the ADS-B display and Wing interface nearly exclusively, and in ways that demonstrated they were gathering pertinent information relevant to their job
duties and were engaged in the tasks. These interactions were largely consistent across tasks and trials in nature and what
the PICs interacted with, although the rate of interactions were affected by unexpected events. 

Overall, none of the experiment's conditions had a negative impact on the PIC's performance. Generally, the PICs performed their job duties successfully and generally the same when the number of nests and number of active UAS increased, as well as during the DAA encounters and adverse weather conditions. 
\section{Acknowledgement}

The reported evaluation activities were supported through University Space Research Association award 904186092. The contents are those of the authors and do not represent the official views of, nor an endorsement, by the National Aeronautics and Space Administration
or the U.S. Government.

The authors thank Wing, their Wing collaborators, and Wing's pilots for supporting the Phase I evaluation. 
\clearpage

\phantomsection
\renewcommand{\refname}{Bibliography}
\renewcommand{\thepage}{\arabic{page}}
\addcontentsline{toc}{section}{Bibliography}

\bibliographystyle{ieeetr}
\bibliography{output}
\end{document}